\documentclass[11pt]{article}

\usepackage{natbib}

\usepackage{booktabs}
\usepackage{amsmath}
\usepackage{graphicx}
\usepackage[colorinlistoftodos]{todonotes}
\usepackage[colorlinks=true, allcolors=black]{hyperref}
\usepackage{amsmath, amssymb,amsthm,bm,bbm,fullpage}
\usepackage{mathrsfs}
\usepackage{cite}
\usepackage{colortbl}
\usepackage[all]{xy}
\usepackage{epsfig}
\usepackage{subfigure}
\usepackage{graphics}
\usepackage{multirow}
\usepackage[affil-it]{authblk}
\usepackage{lineno}
\usepackage{setspace}
\usepackage[font=footnotesize]{caption}
\usepackage{lineno}

\theoremstyle{remark} 
	
\providecommand{\keywords}[1]
{
  \small	
  \textbf{\textit{Keywords---}} #1
}

\newenvironment{sciabstract}{%
\begin{quote}\bf}
{\end{quote}}

\usepackage{xspace}

\title{Strategies for preventing and reversing polarized online discourse}

\author[1,2]{Leon Klingborg}
\author[2]{Kenneth Mavor}
\author[3,4,*]{Alexander J. Stewart}

\affil[1]{School of Mathematics and Statistics, University of St Andrews}
\affil[2]{School of Psychology and Neuroscience, University of St Andrew}
\affil[3]{Observatory on Social Media, Indiana University}
\affil[4]{Luddy School of Informatics, Computing, and Engineering, Indiana University}
\affil[*]{stewalex@iu.edu}

\begin{document}

\date{}
\maketitle

\begin{singlespace}
\begin{sciabstract}
Political polarization poses a variety of challenges for modern democracies. Entrenched disagreements on policy can prevent constructive discourse and compromise, and high levels of affective polarization threaten to undermine social cohesion and support for institutions. Finding ways to promote constructive discourse while maintaining free expression has proved a challenge for social media platforms, media outlets and policy makers alike. Here we develop a computational model -- based in psychology -- of online discourse and opinion dynamics under complex individual identities, which we use to assess the capacity of realistic interventions to prevent or reverse polarization. We show that changes to the range of acceptable opinions in a society -- i.e. the Overton window -- have a limited impact on polarization, and that attempts to ``optimize'' the Overton window can even trigger the onset of polarization. In contrast, interventions that shift attention towards under-discussed topics, or increase the costs of violating existing norms, are often effective at preventing polarization, but are less successful at reversing it. Most strikingly, increasing the salience of influential individuals, who model non-polarized discourse, can be highly effective at both preventing and reversing polarization. However we also find that once polarization has set in, even the most successful interventions result in latent extremism when identities are complex. Our work suggests that restricting speech by shrinking the range of acceptable discourse is an ineffective way to tackle polarization, whereas enforcement of existing norms, attention nudges and the presence of elites who model good behavior can be highly effective.
\end{sciabstract}
\end{singlespace}


\clearpage

Political polarization has been studied for decades by Political Scientists, Psychologists and social scientists more generally \citep{Mason:2018,Mason:2015,McCartyShor16,McCarty16,Iyengar:2012,doi:10.1146/annurev-polisci-051117-073034,annurev:/content/journals/10.1146/annurev-polisci-051117-073034}. However, in recent years the growth of contentious and extreme political discourse, and the rise of populist movements across the globe, has made understanding the causes and consequences of polarization more urgent, and has brought the topic to the attention of a wider range of researchers. One consequence of this has been a proliferation of mathematical and computational models, often based in opinion dynamics \citep{doi:10.1073/pnas.2116950118,7900329}, which seek to explain polarization, as well as to explore ways to prevent it. While this work has produced important insights, there is often a disconnect between the assumptions of such models and the empirical and conceptual understanding, developed by social scientists over many years, of the forces that shape polarization. 


In their earliest forms, opinion dynamics models made highly simplified assumptions about the psychology of agents. Individuals were typically represented as holding a single continuous attitude, which was updated through repeated interactions with others in a network. In classic formulations, such as DeGroot-type models \citep{degroot_reaching_1974,Abelson64,7900329}, attitude change was governed by simple averaging processes, leading in many cases to convergence or consensus. Subsequent models introduced bounded confidence mechanisms, in which agents interact only with sufficiently similar others, producing clustering and, under some conditions, polarization. Later extensions incorporated both attraction and repulsion dynamics, such that agents move closer to similar others while distancing themselves from those who are sufficiently dissimilar \citep{Lerman_njp,10.1093/pnasnexus/pgaf082,10.1093/pnasnexus/pgaf184}. These models require some assessment of similarity, which is often operationalized either through distance in opinion space or through fixed group memberships. However, especially in contemporary social media environments, where group boundaries are fluid and often inferred from interaction rather than given a priori, such similarity judgments are likely to be made dynamically rather than on the basis of stable categories \citep{https://doi.org/10.1002/ejsp.334,doi:10.1177/1088868309341563}.

While most models operate on a single attitudinal dimension, some work has begun to explore multidimensional opinion spaces \citep{doi:10.1073/pnas.2102139118}, and  demonstrate how polarization can emerge across multiple interacting dimensions. However even where multiple attitude dimensions are accounted for agents are typically characterized by a unitary internal state, rather than a more complex or context-dependent representation of the self (although see \citep{dalege_networks_2025,zimmaro2025meta}). Another common simplifying assumption in early models is that agents are memoryless, with each interaction treated as independent of prior exchanges. More recent work has begun to relax this assumption by incorporating limited forms of memory or history dependence -- for example, models in which agents sample from previously encountered opinions or where past interactions influence current behavior \citep{brown_social_2022}. These developments represent an important step towards greater psychological realism, as they acknowledge that social interaction is embedded in a temporal context rather than consisting of isolated encounters.

In this paper we build on these developments by introducing several additional psychological elements designed to expand the realism of computational models of polarization. In particular we allow for the possibility of multiple selves (or self-aspects) within each agent. Rather than treating individuals as possessing a single, stable attitudinal position, we allow agents to maintain multiple identity-relevant self-aspects that may or may not align across different issue domains. This reflects the increasingly recognized role of social media as a space for the construction and negotiation of identity, where individuals can express different facets of the self across contexts and interactions \citep{Smaldino2026}. Consequently, agents in our model are not simply holders of multiple opinions, but actors capable of navigating a more complex internal identity structure.

In addition, we incorporate a limited memory of recent interactions, allowing agents to retain attitudes expressed by others in prior exchanges, that can then provide context for subsequent behavior. While interactions on social media are often brief and fragmented, they nevertheless occur within streams of communication that provide a short-term history of who has said what. By modeling this limited memory, we move beyond the assumption of independent interactions and enable agents to evaluate current exchanges in relation to recent experience.

This memory mechanism also allows us to introduce a more psychologically grounded account of attraction and repulsion dynamics. In contrast to many previous models, where a fixed tolerance parameter determines whether an interaction results in convergence or divergence based solely on distance in opinion space, agents in our model use their recent interaction history to assess whether others are relatively similar or dissimilar within the current context. This draws on the logic of the meta-contrast principle from self-categorization theory \citep{turner1987rediscovering}, to enable a more flexible and context-sensitive basis for social influence, in which responses are shaped not only by absolute differences but by their relation to a dynamically constructed reference frame.

Finally, we incorporate two tolerance-like mechanisms that further structure agent behavior. First, a ``soft Overton window'' constrains the likelihood that agents will express positions far outside the range of recently encountered views, reflecting the tendency for communication to remain anchored within locally salient norms. Second, we implement a ``hard Overton window'', representing the broader range of socially acceptable attitudes as defined by the social media platform. This concept parallels longstanding ideas in psychology, such as latitudes of acceptance and rejection, which emphasize that individuals are more receptive to positions that fall within perceived bounds of acceptability. By embedding these mechanisms within the model, we link micro-level interaction processes to dynamically evolving normative constraints. Together, these extensions move beyond the representation of agents as simple, memoryless holders of single attitudes, and instead model them as context-sensitive actors with structured identities and temporally grounded interaction histories. In doing so, the present approach aims to provide a more psychologically realistic account of how polarization emerges and evolves in contemporary social environments.

We use our computational model to classify the dynamics that emerge in populations of interlocutors engaging in thread-like conversations that reflect the structure of conversations on a range of social media platforms such as Reddit, Facebook and X among many others. We systematically vary the psychological parameters of the model in order to characterize the circumstances under which extreme polarized discourse is likely to arise \citep{Bliuc24}. Most importantly, we adopt the approach of resent research on combating misinformation \citep{Coleman22}, and use our large-scale simulations to explore the capacity of realistic interventions, implemented at the level of social media platform rules and algorithms, to prevent or reverse extreme polarization.

\section*{Model \& Results}

We develop a model, grounded in psychology, for attitude dynamics in groups of individuals who possess multi-dimensional identities, and who engage in discourse across multiple topics. We adopt many of the standard assumptions of opinion dynamics models \citep{7900329}, namely that attitudes are shaped by individual preferences for in-group conformity (in-group pull), as well as a desire for out-group differentiation (out-group push). A number of studies have shown that these two forces are sufficient to generate polarized attitude distributions \citep{10.1093/pnasnexus/pgaf082,doi:10.1073/pnas.2102149118}. We introduce a number of extensions to this basic framework, in order to assess the efficacy of interventions aimed at preventing or reversing polarization. In addition to allowing for multi-dimensional identities and discussions across multiple topics, we allow for dynamic, individual-level perceptions of in-group and out-group membership, endogenous and exogenous norm enforcement, and variable levels of attention to different topics. We assume that individuals update their attitudes through a process of noisy, myopic learning, under which they seek to maximize a utility function, which balances the different social pressures captured by our model (Figure 1). The details of this process are given below.

\begin{figure}
    \centering
    \includegraphics[width=0.67\linewidth]{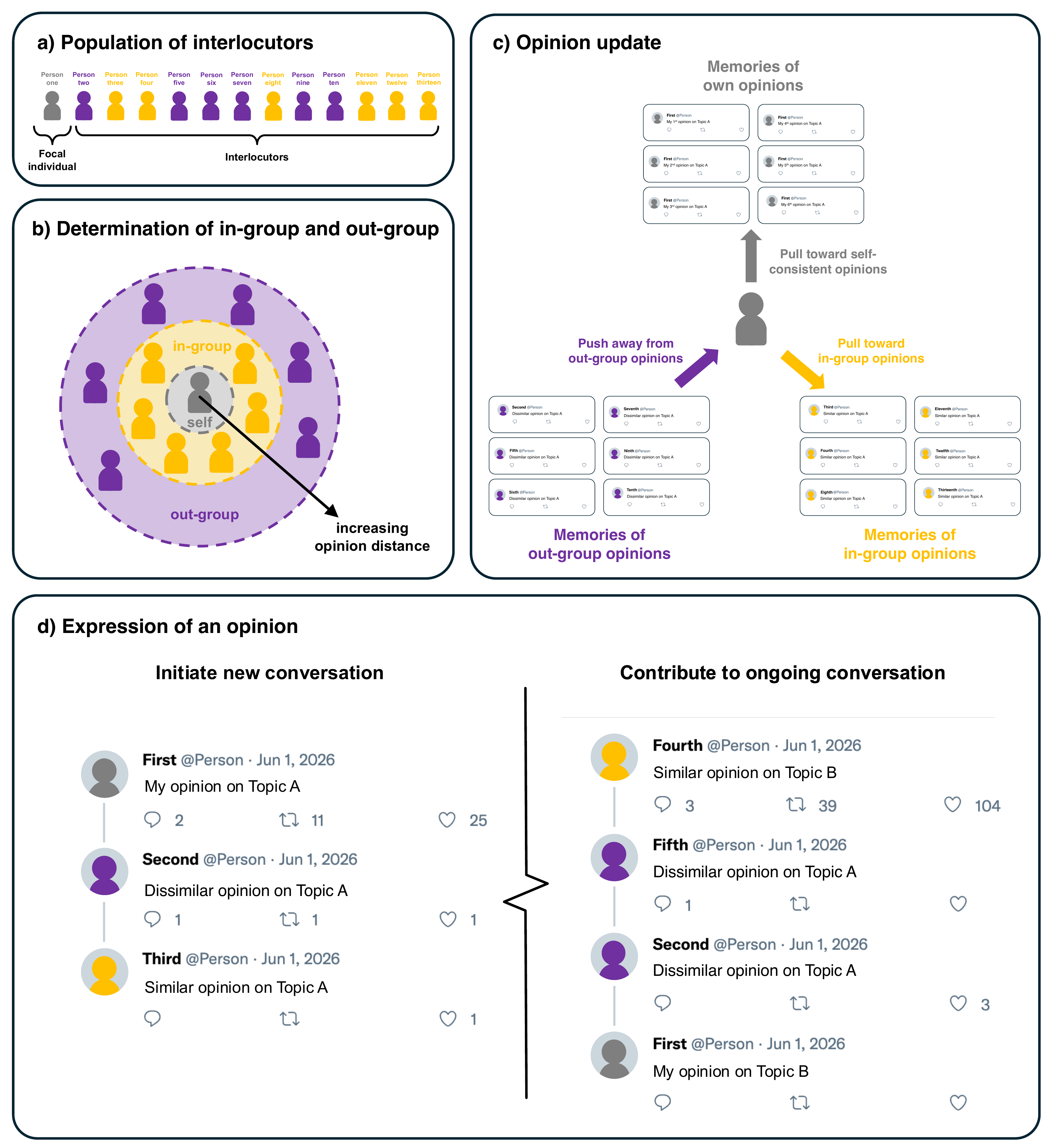}
    \caption{\textbf{Structure of the model.} We model the evolution of online discourse using a computational model that unfolds of a series of stages. a) A focal individual (gray) exists as part of a population of interlocutors, which have previously been divided into an in-group (yellow) and an out-group purple. b) At each step of the model, each individual re-allocates their in-group and out-group based on their memory of previously expressed attitudes/opinions. In particular, they assign those who are closer than average to their own attitude as their in-group, and the rest to their out-group. This is done separately for each identity dimension and each attitude dimension (see main text and Methods). c) Having assigned in-groups and out-groups, each individual decides whether to express an attitude on an ongoing conversation thread, or if no unseen threads are available, to start a new thread on a new topic. They then update their attitude based on the utility of their current attitude and a randomly drawn perturbed attitude (see main text). Each individual's utility balances the desire to conform to their in-group, differentiate themselves from their out-group, and to remain consistent with their previously expressed attitudes. d) Having decided which attitude to express, the focal individual ``posts'' a comment either as the continuation of a thread (on the same topic as previous comments) or else starts a new thread on a topic chosen according to their opinion bias (see main text).}
    \label{fig:1}
\end{figure}

\subsection*{Single Identity Model}

We first describe the dynamics of our model for a population of $N$ individuals, who possess a one-dimensional identity. We then generalize the model to multiple identities in the next section.

Each individual $i$ is characterized by an attitude vector $A^i$ which contains their current attitude on each of $n$ different topics (see Methods). Typically we set $n=2$, so that a group discusses only two topics, however we explore the effect of setting $n>2$ in the SI (section 2.2). Each individual is also characterized by two memory vectors. The first is called the self-memory vector, $S^i$, which contains the $s$ most recent memories of the focal individual's own expressed attitudes. The second is called the other-memory vector, $M^i$, and contains the $m$ most recent memories of other individuals' expressed attitudes. Both memory vectors are of fixed length and are updated sequentially, so that when a new attitude is expressed, this is remembered and the oldest memory the focal individual possess is lost.  We assume that the other-memory vector $M^i$ contains only attitudes that the focal individual has directly encountered (see Methods).

The dynamics of attitude expression then unfold in three stages: i) A individual identifies their in-group based on their current memories of expressed attitudes, ii) they decide whether to express an attitude (either adding to an existing conversation thread or starting a new one) and iii) they decide what attitude to express. 

\subsection*{In-group assignment}
In order to determine their in-group, a focal individual $i$ calculates the average distance $\bar{d}_i$ in attitude space between their own attitude and the attitudes expressed by all other members of the population held in their memory, $M^i$, across all attitude dimensions. We then calculate the average distance of each other individual $k$ from the focal individual, $d_{ik}$ and assign all individuals $k$ for who $d_{ik}\leq \bar{d}_i$ to be the in-group of $i$, and all other individuals to the out-group. This reflects the intuition that individuals try to balance the desire to differentiate from an out-group while conforming to an in-group. This also differs from approaches that apply ingroup/outgroup status on the basis of some stable characteristic and reflects instead the idea that within an opinion-space, groups are formed on the basis of shared opinions as much as by shared characteristics. 
And so, our method of in-group assignment is dynamic, with the size of the in-group and out-group depending both on the memories of the focal individual, and the distribution of previously encountered attitudes.

 \begin{figure}
    \centering
    \includegraphics[width=0.8\linewidth]{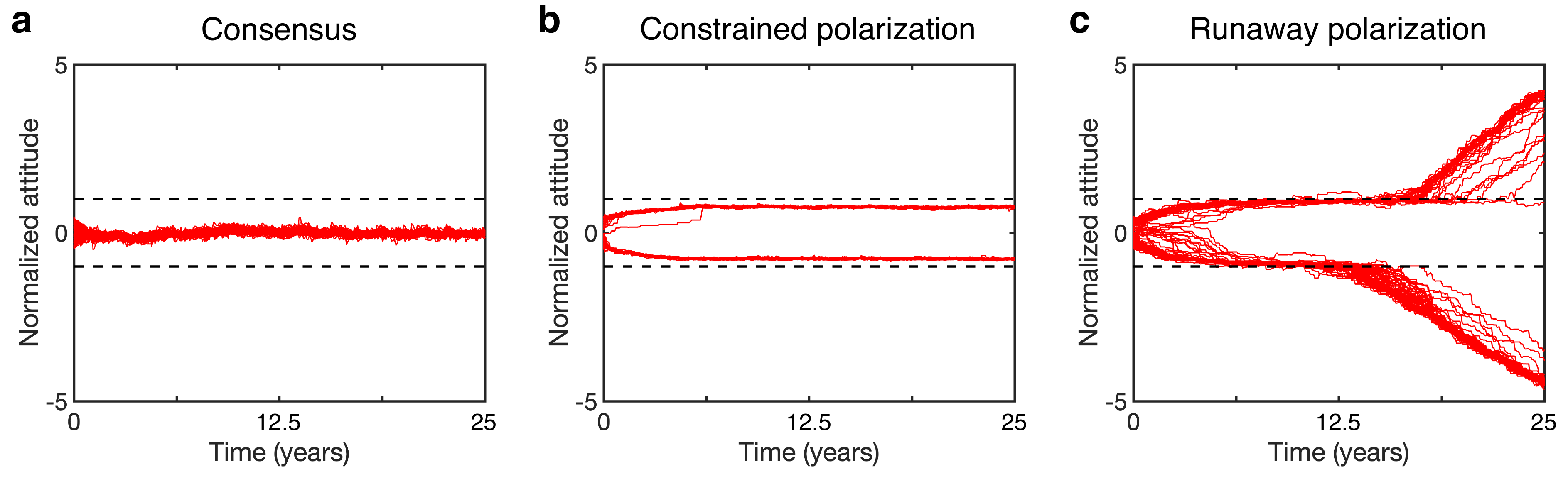}
    \caption{\textbf{Classification of states}. Panels a, b, and c show examples of the three qualitatively different outcome states of our model and the timeline of how they evolve of the course of twenty years, based on a posting rate of 5 posts per day per individual. Each red line represents the attitude associated with one identity dimension belonging to one individual. a) Attitudes converge resulting in a broad consensus, without any clear separation into distinct clusters associated with polarization. b) Attitudes break into distinct two clusters, but remain constrained by the hard Overton window (dashed black lines), corresponding to constrained polarization. c) Attitudes break into two distinct clusters, and then escape the hard Overton window, corresponding to runaway polarization. Simulations were run for the equivalent of 25 years (50,000 individual attitude updates) with $N=100$ individuals, $n=2$ attitude dimensions, $D=2$ identity dimensions, $m=50$ memories of others, $s=25$ self-memories and relative attitude update step size of $r=0.2$. Utility parameters consist of self weight  $\lambda_{\text{self}}=0.4$, an out-group pull of $\lambda_{\text{pull}}=0.4$, soft Overton cost of $c_S=0.5$ and hard Overton cost of $c_H=0.5$. The consensus state example in panel (a) has an attention bias of $\alpha=0.0$ and Overton window width of $w_H=2.5$. The constrained polarization state example in panel (b) has attention bias of $\alpha=0.45$ and Overton window width of $w_H=10.0$. The runaway polarization state example in panel (c) has attention bias set to $\alpha=0.9$ and Overton window width at $w_H=10.0$. Attitudes are normalized with respect to the Overton window width for ease of comparison.}
    \label{fig:2}
\end{figure}

\subsection*{Deciding whether to express an attitude}
Once all focal individuals have assigned their in-group, they next decide whether to express an attitude. For each individual in the population, if there are any posts that they have ``not seen'' (i.e. a post that they have not yet encountered and decided whether to respond to) they choose an ``unseen'' post at random, and respond with probability $p(r)=\frac{1}{N+r}$ where $r$ is the ``depth'' of the thread -- i.e. the number of replies between the original post and the current post being considered. If the post being considered is the original post, $r=0$. If the focal individual decides not to respond to the post, it is considered ``seen'' and it is not revisited by that individual again, nor is it committed to memory.

This ensures i) that the expected number of replies to a given initial post is 1 and ii) that replies become less likely as the depth of the thread increases (see SI section 1). The result is an exponential distribution of thread lengths -- we do not model the types of ``mega-threads'' seen on platforms such as Reddit or X, which result in a more right-skewed distribution \citep{Avalle24}.  All replies are assumed to be on the same topic as the post they are responding to, meaning that all threads are on a single topic (see Figure 1). If there are no unseen posts to respond to, the focal individual starts a new thread. In particular, they start the new thread on topic 1 with probability $\beta$, and otherwise select topic 2. For simplicity we set $0.5\leq\beta\leq 1$, i.e. topic 1 is always considered the ``dominant'' topic. We then define ``attention bias'' to be $\alpha=2\beta-1$, so that attention bias is 0 when $\beta=0.5$ (both topics discussed with equal probability) and attention bias is 1 when $\beta=1$ (only topic 1 is discussed). We may consider attention bias as reflecting to some extent the impact of algorithmic decisions on the part of social medial platforms, such as algorithmic news feeds which emphasize certain topics or types of post at the expense of others.
Note that we consider the case of 3 or more topics in SI Section 2.2.

\subsection*{Expressing an attitude}
When a focal individual $i$ has decided to post a reply or start a thread on a topic $j$, they can choose to either express their current attitude denoted, $a^i_j$, or update their attitude in a manner that increases their expected utility. The expected utility of an attitude $a^i_j$ is given by

\begin{equation}
    U(a^i_j)=(1-\lambda_{\text{self}})\left(\,
   (1-\lambda_{\text{pull}})d^j_{\text{out}}-
    \lambda_{\text{pull}}d^j_{\text{in}}
    \right)-d^j_{\text{self}}\lambda_{\text{self}}    
\end{equation}

where $\lambda_{\text{self}}$ is the weight given to the desire for self-consistency vs the strength of the desire to conform to social pressure and $\lambda_{\text{pull}}$ is the weight given to ``in-group pull'' vs ``out-group push'', i.e. the desire to balance differentiation from the out-group and conformity to the in-group. These parameters are assumed to be the same for all members of the population, reflecting a shared set of preferences in response to a common environment. Cases where parameters vary between individuals are considered in the SI Section 2.1. The distance $d^j_{\text{in}}$ is the average distance between the focal individual's attitude on topic $j$ and all members $k$ of their in-group $\mathcal{I}$ contained in their memory $M^i$. Similarly the distance $d^j_{\text{out}}$ is the average distance between the focal individual's attitude on topic $j$ and all members $k$ of their out-group $\mathcal{O}$ contained in their memory $M^i$. Finally the distance $d^j_{\text{self}}$ is the average distance between the focal individual's current attitude on topic $j$ and all of their self-memories $S^i$ (see Methods)

The focal individual chooses an updated attitude $\tilde{a}^i_j=a^i_j+\Delta a$ where $\Delta a$ is a perturbation drawn from a normal distribution with mean 0 and standard deviation 1. They then compare the utility of their current attitude to the utility of the perturbed attitude, i.e. they calculate the utility difference $\Delta A_u=U(a^i_j)-U(a^i_j+\Delta)$. They then express the perturbed attitude, $\tilde{a}^i_j$,  with probability 

$$
p=\frac{1}{1+e^{h_a \Delta A_u}},
$$

and otherwise they express their current attitude $a^i_j$. Here $h_a$ scales the steepness of the sigmoid function, and we set $h_a=10$ by default. If the focal individual chooses to express the perturbed attitude, they also update their current attitude in the direction of the perturbation, i.e. $a^i_j\to a^i_j+r\Delta a$, where $r$ controls the step size in attitude space, and by default we set $r=0.2$. Whatever attitude the individual expresses, this is then added to their self-memory vector $S^i$.

\begin{figure}
    \centering
    \includegraphics[width=0.67\linewidth]{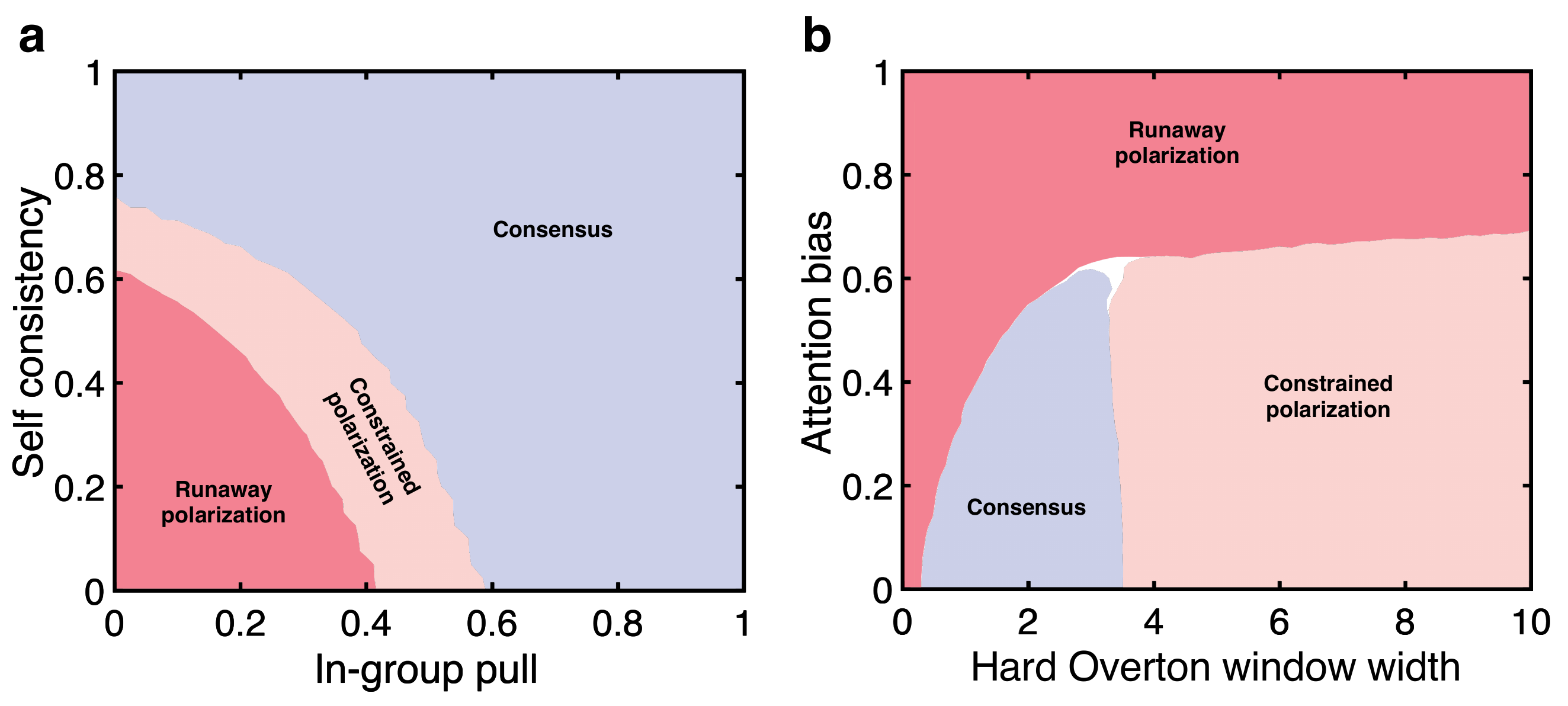}
    \caption{\textbf{Long-run outcome of discourse.} Consensus, constrained polarization and runaway polarization under a range of model parameters. a) We systematically varied the self consistency weight, $\lambda_{\text{self}}$ and the in-group pull weight $\lambda_{\text{pull}}$. We see that both parameters effect the outcome of discourse similarly. When weights are low, runaway polarization takes hold (dark pink). As they increase there is a sharp transition to constrained polarization (light pink) and then to consensus (blue). b) We also systematically varied the hard Overton window, $w_H$, and the attention bias $\alpha$. Here we see that increasing attention bias tends to lead to runaway polarization regardless of the hard Overton window width. However when the window width is small, runaway polarization takes hold, whereas intermediate widths promote consensus and wide windows lead to constrained polarization. Simulations were run for the equivalent of 50 years (10 million attitude updates) with $N=100$ individuals, $n=2$ attitude dimensions, $D=2$ identity dimensions, $m=50$ memories of others, $s=25$ self-memories, attitude update step size of $r=0.2$, soft Overton cost at $c_S=0.5$ and hard Overton cost at $c_H=0.5$. In panel (a) attention bias is $\alpha=0$ and hard Overton window width $w_H=5$. In panel (b) self weight is $\lambda_{\text{self}}=0.4$ and in-group pull is $\lambda_{\text{self}}=0.4$. A given set of parameters are classified to belonging to a particular state if they result in that classification at least 50\% of the time over 200 replicate simulations with initial conditions randomized within the hard Overton window. The white region in panel (b) corresponds to cases where no outcome state meets this threshold.}
    \label{fig:3}
\end{figure}

\subsection*{Generalization to multiple Identities}
 Our model can be generalized to account for complex identities, composed of $D$ different ``identity dimensions'', each associated with a different attitude, and a corresponding in-group. This reflects a scenario in which an individual may hold different attitudes on the same topic in different contexts, e.g., recent work has demonstrated cognitive switching between "parent" and "feminist" identities using linguistic data \citep{https://doi.org/10.1111/bjso.12906}.

 In order to account for this kind of complexity, we assume that identity dimension $\alpha$ of individual $i$ has an associated attitude vector $A^i_\alpha$. The utility of each identity dimension, $U_\alpha$, is calculated according to Eq. 1, and the attitude associated with that identity is expressed with probability

$$
    q_\alpha= \frac
    {\frac{1}{1+e^{-h_u \Delta U_\alpha}}}
    {\sum^{D}_{\gamma=1}\frac{1}{1+e^{-h_u \Delta U_\gamma}}}
$$

where $\Delta_{U_\alpha}$ is the difference between the utility of identity $\alpha$ and the average utility across all $D$ identity dimensions and $h_u$ scales the steepness of the softmax activation function, where we set $h_u=10$ by default. Note that the distances in the utility calculation, as well as the assignment of in-groups and out-groups, are calculated separately for each identity dimension, and that we assume that the memory vectors $M^i$ and $S^i$ do not include any information of the identity dimension associated with the expressed attitude. We set $D=2$ by default, although we also explore the effect of varying the number of identity dimensions on polarization (see Figure 5).

\subsection*{Timescales}
The temporal dynamics of our model are recorded in terms of posts per individual. The precise timescale this corresponds to in reality depends both on how often interlocutors post and on how rapidly they update their attitudes. The rate of social media posting is less frequently measured than overall social media use. However, according to a Morning Consult survey of US adults in 2025, 33\% of social media users post at least daily \citep{MorningConsult} while Pew find that the most active 10\% of X users post $>5$ times per day \citep{Pew}. 

We assume that the population of interlocutors in our model belong to such a group of active users. For simplicity, we assume that each individual make around 5 posts per day, or 2000 posts per year so that in a population of 100 individuals, 200,000 posts are produced each year. This means that the temporal dynamics in our model, which occur over the course of millions of posts, correspond to a timescale of multiple years (Figure 2). While this is slow, it is comparable to the timescales on which changes in affective polarization are observed in the US \citep{annurev:/content/journals/10.1146/annurev-polisci-051117-073034,Iyengar:2012}.

 \subsection*{The Overton window}

 Our model captures the key ``forces'' identified by social psychologists and political scientists, which shape attitude formation over time, while also capturing the basic features of ``thread-like'' online discourse via written text, as seen on platforms like X, Reddit and Facebook, among many others. However we have not so far accounted for the impact of ``moderation'', i.e. formal or informal policies in a given online community about the range of acceptable discourse, sometimes referred to as the ``Overton window''.

 Such moderation occurs in different forms on different social media platforms. We capture its effect in two ways. 
 First we assume that on a given platform, there is a fixed range of acceptable discourse, and when an individual expresses (or contemplates expressing) an attitude outside this range they suffer a cost, $c_H$, which is applied to their utility calculation (Eq. 1). We call this fixed range of acceptable attitudes the ``hard'' Overton window, and we use it to capture the effects of platform enforced speech rules, such as adding tags to misleading posts or deleting posts that express extreme views. Second, we assume that when an individual expresses (or contemplates expressing) an attitude that lies outside the attitudes held by the population (taken collectively), they suffer a cost $c_S$ to their utility. We call this flexible range of acceptable attitudes the ``soft'' Overton window, and we use it to capture the effects of emergent community enforced speech rules, such as down-voting a particular comment (see Methods for full details).

\begin{figure}
    \centering
    \includegraphics[width=0.5\linewidth]{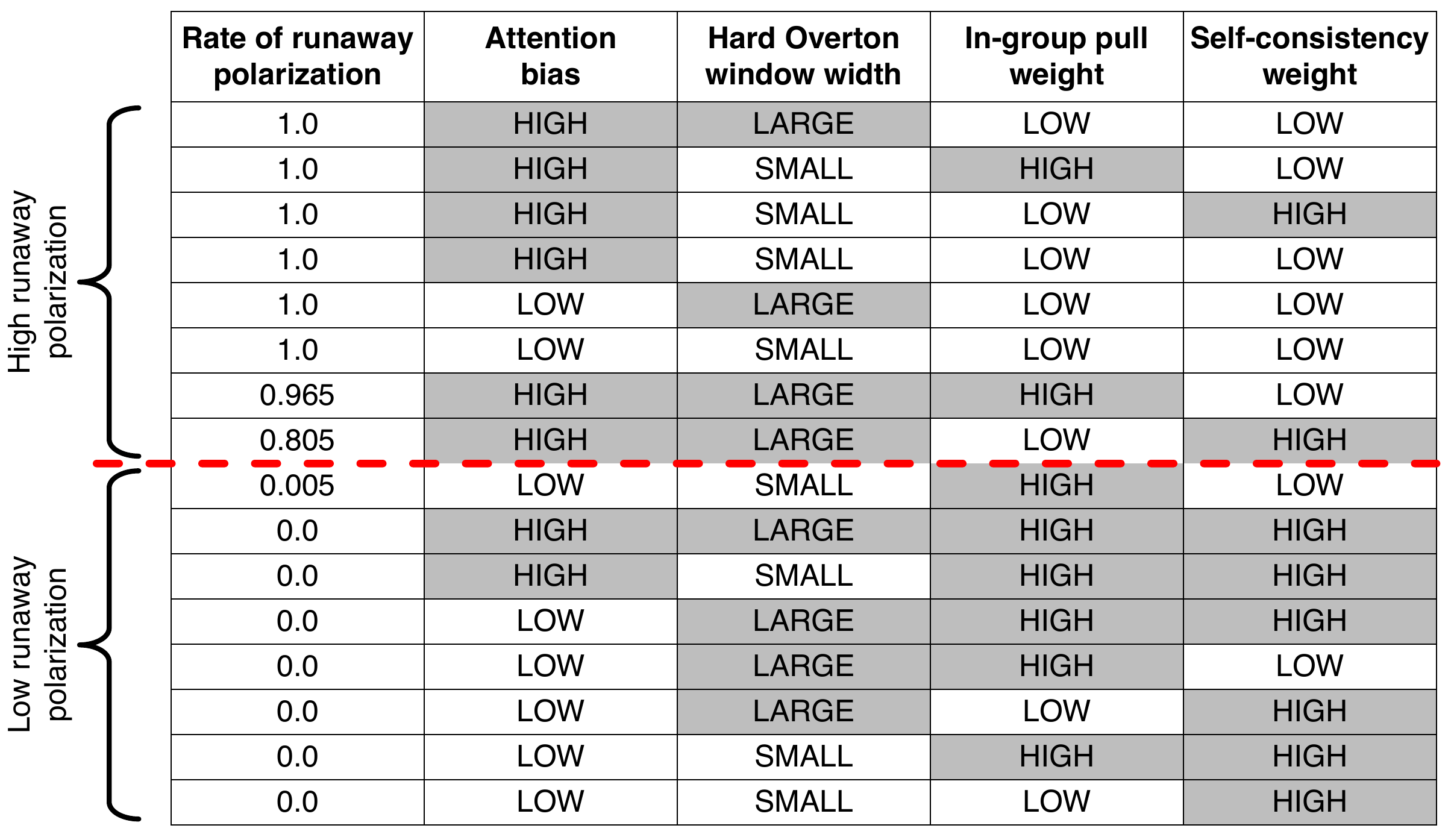}
    \caption*{\textbf{Table 1. Combinations of parameters.} We explored further combinations of  the four parameters varied in Figure 3. To do this we set: Attention bias $\alpha$ to either ``HIGH'' ($\alpha=0.75$ or ``LOW'' ($\alpha=0$); Hard Overton window width $w_H$ to either ``LARGE'' ($w_H=5$ or ``SMALL'' ($w_H=1$); In-group pull weight $\lambda_{\text{pull}}$ to either ``HIGH'' ($\lambda_{\text{pull}}=0.6$ or ``LOW'' ($\lambda_{\text{pull}}=0.2$); Self-cosistency weight $\lambda_{\text{self}}$ to either ``HIGH'' ($\lambda_{\text{self}}=0.6$ or ``LOW'' ($\lambda_{\text{self}}=0.2$). These parameter values were chosen based on the outcome states observed in Figure 3. For each of these 16 combinations we calculated the proportion of simulations that resulted in runaway polarization. All other parameters are the same as described for Figure 3.}
    \label{Tab:1}
\end{figure}

\subsection*{Classification of observed discourse}

In order to study the dynamics of polarization in our model, we simulated populations of $N=100$ individuals engaging in discourse over a number of updates that correspond to between 20 and 50 years, as described above. At the end of this period we systematically classified (see Methods) the state of the population in terms of the distribution of attitudes. In particular we determined i) whether the distribution of attitudes had broken into two or more distinct clusters and ii) whether attitudes remained constrained by the hard Overton window (see Figure 2).

This results in three qualitatively different observed states: i) Discourse remains unpolarized and remains constrained by the hard Overton window. We call this ``consensus''. ii) Discourse becomes polarized, but remains constrained by the hard Overton window. We call this ``constrained polarization''. iii) Discourse becomes polarized and is not constrained by the hard Overton window. We call this ``runaway polarization''. We do not observe instances of other types of state (e.g. consensus that nonetheless escapes the hard Overton window. See SI Section 1). Of these three states, we treat the third, runaway polarization as an unambiguously negative outcome, since it corresponds to polarized discourse that also violates the rules set by the platform, and therefore can be interpreted as corresponding to ``extreme'' speech.  We systematically explore the conditions under which different outcome states are favored, as well as simulated interventions aimed at preventing or reversing runaway polarization.

\subsection*{Internal preferences and the external environment}
We begin by systematically varying four key parameters of the model (Figure 3 and Table 1) namely: the self-consistency weight, $\lambda_{\text{self}}$; the in-group pull weight $\lambda_{\text{pull}}$; the degree of attention bias, $\alpha$ and the width of the hard Overton window, $w_H$. Other combinations of parameters are explored in the SI Section 2. We find that increasing the self-consistency weight and increasing the in-group pull have qualitatively similar impact on the expected outcome of discourse dynamics, with decreasing values of both (i.e. increasing levels of social pressure or out-group push) leading to a transition from consensus, to constrained polarization and then to runaway polarization (Figure 2a). The interaction between attention bias and hard Overton window width, by contrast, is more complex (Figure 2b). We find that, broadly speaking, an increase in the hard Overton window width results in a transition away from consensus and towards constrained polarization, but that this alone does not lead to runaway polarization. However a very narrow hard Overton window can lead to runaway polarization, since discourse trajectories easily escape by chance (Figure 2b). Increasing attention bias on the other hand leads to a transition away from either consensus or constrained polarization, towards runaway polarization. Taken together, these results show that the outcome of discourse is sensitive both to the internal preferences of individuals ($\lambda_{\text{self}}$ and $\lambda_{\text{pull}}$) and to the external environment shaped by the platform ($\alpha$ and $w_H$). 

We further explore the interaction between these parameters in Table 1, in which we characterize the impact of 16 conditions, in which attention bias, $\alpha$, self-consistency weight, $\lambda_{\text{self}}$, and in-group pull weight, $\lambda_{\text{pull}}$, are set to either ``HIGH'' or ``LOW'' values, and the hard Overton window width, $w_H$ is set to either a ``LARGE'' or ``SMALL'' value. We see that runaway polarization occurs at high rates when at least two of the three following conditions are met: i) High attention bias ii) Low self-consistency weight iii) Low in-group pull (Table 1). In the SI Figure S14-S15 and Table S1 we perform similar analyses for the costs of violating the soft and hard Overton windows, $c_S$ and $c_H$, as well as for over 1000 combinations of the six parameters discussed above.

\begin{figure}[h]
    \centering
    \includegraphics[width=0.67\linewidth]{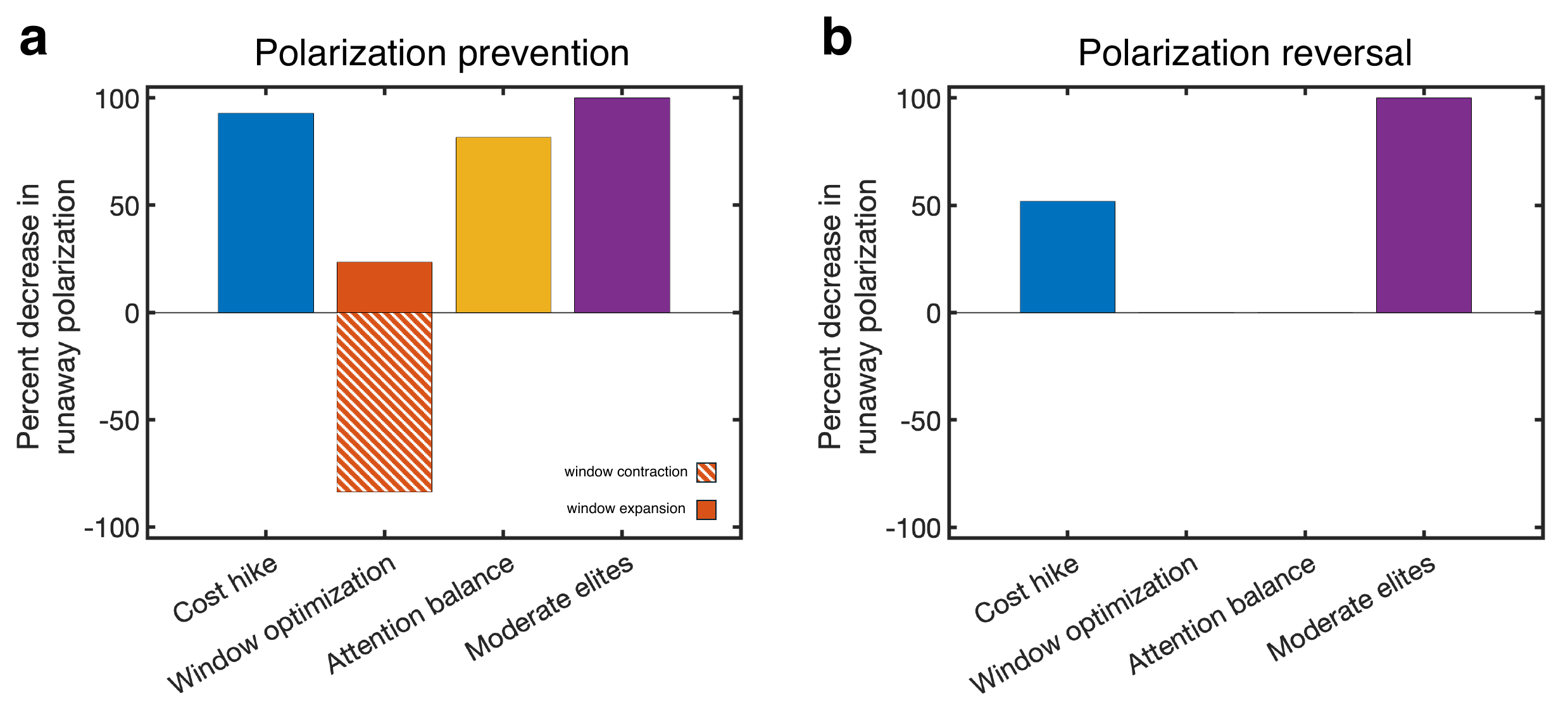}
    \caption{\textbf{Intervention efficacy}. We calculated the average efficacy of the four interventions described in the main text across hard Overton window widths varying from 0.05 to 10 and attention bias varying from 0 to 1. a) We calculated the percentage decrease in runaway polarization with and without the intervention. In the case of the window optimization intervention, we broke the result down into interventions that decrease the hard Overton window (striped red) and interventions that increase the hard Overton window (solid red). This intervention is largely ineffective, and results in a backfire effect when the window width is decreased. All other interventions are broadly effective. b) We also assessed the capacity of each intervention to reverse runaway polarization. We see that only cost hike and moderate elites intervention have any effect, with moderate elites being the most effective.  Simulations are run for the equivalent of 50 years (10 million attitude updates) with interventions implemented after 2.5 years (500 000 attitude updates). Simulations are run with $N=100$ individuals, $n=2$ attitude dimensions, $D=2$ identity dimensions, $m=50$ memories of others, $s=25$ self-memories and attitude update step size of $r=0.2$, self weight is $\lambda_{\text{self}}=0.4$ and in-group pull is $\lambda_{\text{self}}=0.4$. Soft Overton cost is $c_S=0.5$ and hard Overton cost is $c_H=0.5$.  The cost hike intervention hikes the cost of transgressing the Overton window to $c_H=50$ (100 fold), the window optimization intervention sets the Overton window width to 3.0, and the weight of the ingroup pull of moderate elites is set to 1.0 for the moderate elites intervention (see Methods). The attention balance intervention sets $\alpha=0$.}
    \label{fig:4}
\end{figure}

\subsection*{Interventions}

We next explore a set of four interventions, aimed at preventing or reversing runaway polarization. To do this we specify four changes, which correspond to the types of interventions that are readily available to social media platforms. These interventions are

\begin{enumerate}
    \item Increasing the cost of transgressing the hard Overton window, $c_H$. This corresponds to stronger penalties for violating platform set moderation rules.
    \item Optimizing the hard Overton window. This is done by changing the width of the hard Overton window to the value that produces the maximum consensus (see Figure 2b) and corresponds to relaxing or tightening rules around acceptable discourse.
    \item Balancing attention across topics (i.e. setting $\alpha=0$). This corresponds to using algorithmic news feeds to amplify under-discussed topics \citep{pennycook_shifting_2021}.
    \item Introducing moderate elites who exert a ``pull' towards the center of the hard Overton window. This corresponds to amplifying moderate but influential voices, again via algorithmic news feeds or similar approaches \citep{BETTS2026103179,10.1093/pnasnexus/pgae224,https://doi.org/10.1155/2018/2740959,BACK2023102639}.
\end{enumerate}

We first implement these interventions separately at a fixed time (2 years) after the start of the simulation. We explore the average decrease in runaway polarization across a range of initial hard Overton window widths, and levels of attention bias, and compare the outcome state of the discourse after the intervention to the expected outcome state when the intervention is not implemented (Figure 4a). We see that interventions 1, 3 and 4 are highly effective (80-100\%) at preventing runaway polarization across the parameters explored. Intervention 2 (optimizing the hard Overton window width) by contrast is largely ineffective. In particular, when optimization corresponds to expanding the window width, there is prevention in 23\% of cases. However when optimization corresponds to contracting the hard Overton window we observe a strong backfire effect, with runaway polarization increasing by 83\% compared to the baseline condition.

\begin{figure}[h]
    \centering
    \includegraphics[width=0.6\linewidth]{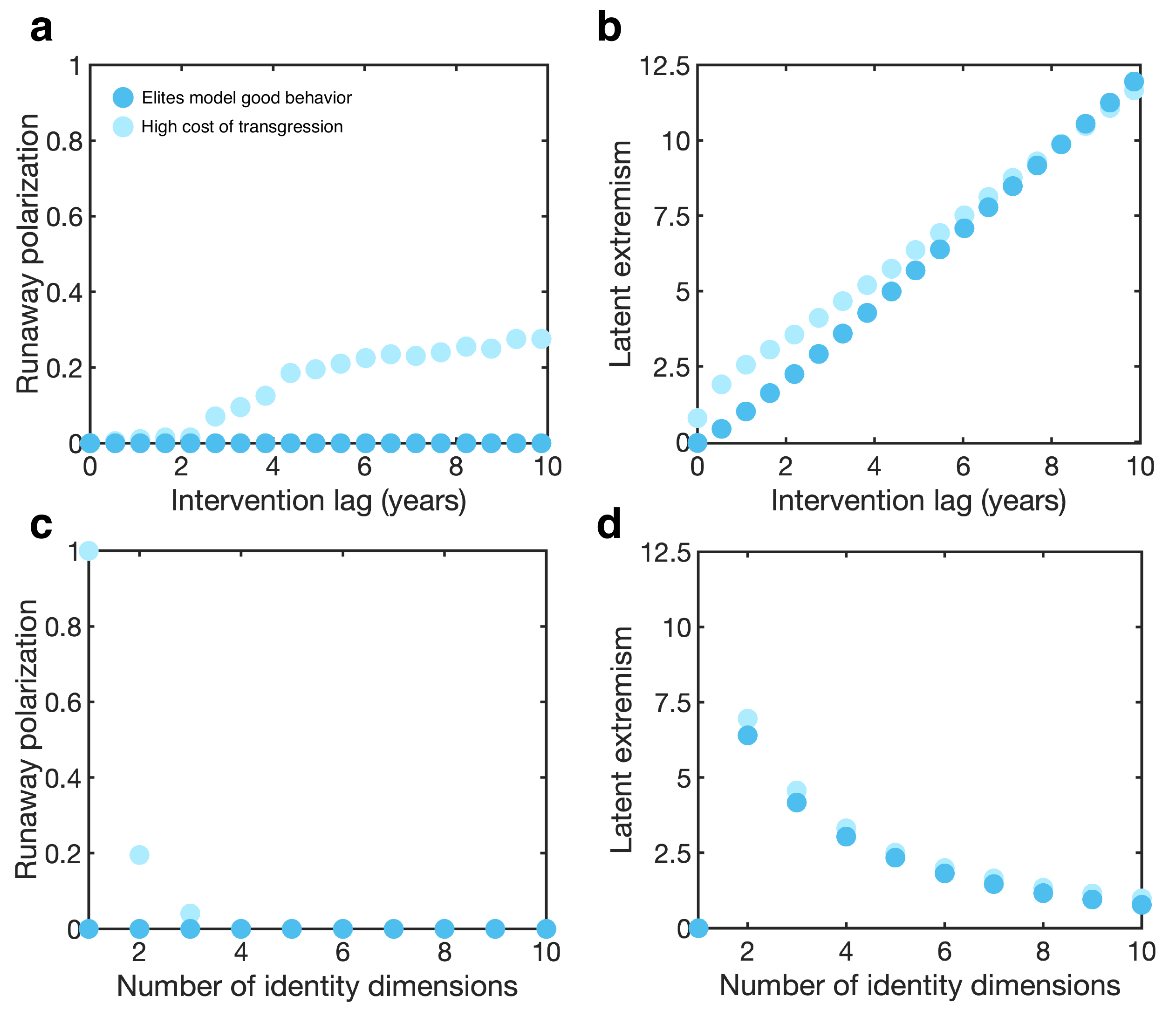}
    \caption{\textbf{Responsive interventions.} We implemented the two interventions observed to reverse runaway polarization in Figure 3 (cost hike and moderate elites) and varied the time, after the onset of runaway polarization, at which interventions were implemented, as well as the number of identity dimensions $D$. a) The cost hike intervention (light blue) is moderately sensitive to the time lag of the responsive intervention, where as the moderate elites intervention is always effective (dark blue). b) Both interventions result in latent extremism, which grows with the lag of the responsive intervention. c) Similarly, the cost hike intervention is sensitive to the number of identity dimensions, whereas the moderate elites intervention is always effective. d) Both interventions display similar levels of latent extremism (note latent extremism is by definition 0 when $D=1$). The average latent extremism across identities declines as the number of identity dimensions decreases for values $D\geq 2$. This reflects the fact that typically only a single identity dimensions experiences runaway polarization in the first place. Simulations were run for the equivalent of 50 years (10 million attitude updates) with $N=100$ individuals, $n=2$ attitude dimensions, $m=50$ memories of others, $s=25$ self-memories and attitude update step size of $r=0.2$. Self weight is $\lambda_{\text{self}}=0.4$ and in-group pull is $\lambda_{\text{self}}=0.4$. Soft Overton cost is $c_S=0.5$, hard Overton cost is $c_H=0.5$, hard Overton window width is $w_H=1.0$, and attention bias is $\alpha=0.75$. The cost hike intervention makes the cost of transgressing the Overton window 100 times greater and the weight of the ingroup pull of moderate elites is set to 1.0 for the moderate elites intervention. In panels a and b the number of identity dimensions is set to 2 and in panel c and d the intervention lag is set to 4 years (1 million attitude updates).}
    \label{fig:5}
\end{figure}

Next we explored the capacity of the four interventions to reverse runaway polarization in the cases where it has already set in at the time of the intervention (Figure 4b). We find that only interventions 2 and 4 have any effect on reversing polarization, with intervention 2 (increasing the cost of transgressing the hard Overton window) reversing runaway polarization in 52\% of cases, while intervention 4 (amplifying moderate elites) being successful in all cases across the parameters explored. We further find (SI Fiugre S19) that the efficacy of a moderate elites intervention displays threshold-like behavior, becoming effective when the strength of the elite pull exceeds a certain value (around 0.2, see Figure S19). Taken together these results suggest that attention nudges, increasing costs for extreme posts and amplifying moderate elites can all be effective at preventing runaway polarization, while cost hikes and moderate elites are the only interventions among those considered that can reverse it. Of these, moderate elites is clearly the most effective.

\subsection*{Responsive interventions and latent extremism}

When considering the capacity of different interventions to reverse runaway polarization, it is realistic to consider a ``responsive'' scenario, in which the intervention is applied at a fixed time after its onset (rather than at a fixed time after the beginning of the simulation, as in Figure 4). We explore the effect of such responsive interventions in Figure 5a-b, for the two interventions (1, cost hike and 4, moderate elites) previously found to be effective. We find (Figure 5a) that the efficacy of intervention 1 is somewhat sensitive to the time lag after the onset of runaway polarization, where as intervention 4 is effective regardless of the lag for the parameters explored (see also SI Figure S17, S19 and S20). However we also
observe that reversal of runaway polarization typically involves a ``switch'' from expressing more extreme attitudes associated with one identity, to expressing less extreme attitudes associated with another identity. The result is that individuals maintain ``latent extremism'' associated with the under-expressed attitude and corresponding identity dimension. We calculate latent extremism as described in Materials \& Methods below. Qualitatively, we capture the extent to which unexpreessed attitudes tend to lie outside of the hard Overton window. Latent extremism is associated with both successful interventions, and increases linearly with the time lag of the intervention (Figure 5b).

Finally we explored the impact of the number of identity dimensions on both the efficacy of responsive interventions at reversing runaway polarization, and on the average degree of latent extremism across all identity dimensions, $D$ (Figure 5c-d). We find that the cost hike intervention (1) is sensitive to the number of identity dimensions, and is ineffective when $D=1$, but is fully effective for the parameters considered once $D>3$. The moderate elites intervention (4) on the other hand is insensitive to the number of identity dimensions and is always effective for the parameters considered (Figure 5c). We observe that latent extremism is similar for both interventions, and tends to decline as the number of identity dimensions increases, reflecting the fact that only the attitudes associated with a single identity dimension tend to experience runaway polarization (Figure 5d, also see SI Figure S18-S20).

\subsection*{Extensions and robustness checks}

Our results are purely computational in nature and involve parameters, such as utility weights, that are difficult to estimate empirically. And so in the SI we have systematically varied the model parameters discussed in the main text, in order to assess the robustness of our findings. We group the parameters broadly into three categories: i) Utility parameters. ii) Cognitive parameters. iii) Intervention parameters (see SI section 2). 

We find (SI section 2.1) that heterogeneity in the key utility parameters, self-weight, $\lambda_{\text{self}}$, and in-group pull, $\lambda_{\text{self}}$ can lead to runaway polarization, as the more susceptible individuals trigger polarization in the wider population (Figure S6) \citep{doi:10.1073/pnas.2102140118,doi:10.1126/sciadv.abd4201}. We also see a similar pattern when we allow for heterogeneity in sensitivity to exogenous costs (Figure S7).

We find (SI section 2.2) that cognitive parameters, such as memory length, $m$ and $s$, number of attitude dimensions, $n$, and number of identity dimensions $D$, have disparate effects on runaway polarization. Longer memories tend to reduce runaway polarization, since they make it easier for individuals to anchor their expressed opinions based on their own past expression, or the expression of other members of the group (Figure S12-S13). A larger number of attitude dimensions tends to make runaway polarization more likely, as this leads to more topics that are under-discussed (Figure S8). More identity dimensions -- i.e. more complex identities -- have little effect on the outcome state of the system absent interventions (Figures S9-S10), but do increase the efficacy of some interventions, most notably the cost hike intervention (Figure 5 and Figures S17-S18) but also the attention balance intervention (Figure S17).

Finally we find that the success of the two most effective interventions -- cost hike and moderate elites -- show threshold-like behavior as we increase the strength of the intervention. Amplifying moderate elites undergoes a very sharp transition, from being completely ineffective to being fully effective, across a wide range of parameters (Figure S19). The cost hike intervention on the other hand displays a less sharp transition as the strength of the intervention is increased (Figure S20).

\section*{Discussion}

Our model is grounded in a set of well-established psychological theories that collectively emphasize the complexity, structure, and context-dependence of the self, as well as the role of social influence in shaping attitudes. Rather than treating individuals as unitary holders of fixed attitudes, these frameworks conceptualize the self as a multifaceted and dynamically activated system, embedded within a social environment.

A central premise of our approach is that individuals possess multiple self-aspects, which may be differentially activated across contexts. This idea is strongly supported by work on self-complexity \citep{linville_self-complexity_1987}, which demonstrates that individuals organise self-knowledge into distinct domains that can vary in their degree of differentiation and overlap. Similarly, the theory of self-schemas \citep{markus_self-schemata_1977} highlights that people hold structured cognitive representations of the self that guide attention, interpretation, and behavior in domain-specific ways. More recently, the Multiple Self-Aspects Framework (MSF) \citep{doi:10.1177/1088868310371101} expands on these ideas by explicitly modeling the self as a network of context-dependent self-aspects that can be selectively activated depending on situational cues. In particular, we show that increasing the number of identity dimensions in our model, which corresponds to increasing the number of self-aspects available to be activated, does not tend to effect the outcome state of the model in isolation (Figures S9-S10) but does make interventions that increase attention for under-discussed topics, and especially interventions that increase the costs of violating the hard Overton window, more effective (Figure 5 and Figure S17-S18).

Together, these perspectives converge on the view that the self is not a single, stable entity, but rather a dynamic configuration of partially independent representations, which may or may not align across different domains. As we show here, this has important implications for modeling opinion dynamics. In real-world settings—particularly in online environments—individuals do not simply hold multiple attitudes; they may express different versions of themselves depending on the audience, topic, or interaction context. Social media platforms, in particular, provide opportunities for the ongoing construction and negotiation of identity, allowing users to foreground different self-aspects across interactions.

In addition to this internal complexity, social identity processes play a central role in shaping how individuals evaluate and respond to others \citep{doi:10.1177/1948550611409245,abelson1982affective}. According to Social Identity Theory and Self-Categorization Theory \citep{turner1987rediscovering,https://doi.org/10.1111/j.1751-9004.2007.00066.x}, individuals define themselves in terms of group memberships, and these categorizations are both context-dependent and relational. When a particular social identity is salient, individuals are motivated to align with ingroup norms and to differentiate from relevant out-groups \citep{10.1098/rsif.2023.0372}. Importantly, group boundaries are not always fixed; rather, they can be inferred dynamically from patterns of similarity and difference in attitudes and behavior. This aligns closely with the requirements of modeling social media environments, where group membership is often ambiguous and must be inferred from interaction.

Building on this contextual account of identity salience, our model draws on the logic of the meta-contrast principle from self-categorization theory. In SCT \citep{turner1987rediscovering}, category boundaries emerge where the ratio of intergroup to intragroup differences is maximized, such that perceived similarity within a category is high relative to differences between categories. Translating this insight into the modeling framework, we infer the boundary between ``similar'' and ``different'' others dynamically from individuals’ recent interaction histories. Specifically, individuals use their limited memory of recent exchanges to construct a local map of attitude space, and evaluate positions in terms of how well they support an optimal differentiation between clusters of viewpoints. In this way, attraction and repulsion are not governed by fixed thresholds alone, but emerge from a context-sensitive process that approximates the meta-contrast logic, allowing agents to flexibly partition their social environment into relatively similar and dissimilar others.

These identity processes are closely linked to mechanisms of social influence. Classic work on latitudes of acceptance and rejection \citep{sherif1961social} demonstrates that individuals are more receptive to messages that fall within a perceived range of acceptable positions, and more resistant—or even oppositional—to positions that fall outside this range. This framework provides a psychological basis for the attraction and repulsion dynamics commonly implemented in opinion dynamics models. However, it also suggests that such dynamics are not governed by fixed thresholds alone, but are shaped by contextual cues, prior exposure, and perceived norms.

Our model incorporates these insights by allowing agents to evaluate others’ positions relative to both their recent interaction history and broader normative constraints. The inclusion of a limited memory of recent interactions reflects the idea that social judgments are made within a temporally bounded context, where recently encountered views shape perceptions of what is typical, acceptable, or extreme. This resonates with research on social sampling and norm perception, which emphasizes that individuals infer social reality from the information available to them in their immediate environment.
Finally, the concept of an Overton window we model captures the idea that there exists a broader, socially shared range of acceptable discourse, which constrains what positions can be publicly expressed or seriously entertained. This notion parallels psychological constructs such as latitudes of acceptance, but operates at a more collective level, reflecting shared understandings of what is considered legitimate or permissible within a given social context.

Taken together, these theoretical perspectives motivate a shift away from models that treat agents as simple, memoryless holders of single attitudes, and toward a representation of individuals as context-sensitive actors with structured, multidimensional selves and socially embedded evaluative processes. By embedding these psychological principles within an agent-based framework, the present model aims to more closely approximate the processes through which identities and attitudes co-evolve in contemporary social environments. A key omission from our model is any explicit role for the material effects of one's opinions (e.g. holding false beliefs about the world may result in harm). Recent work has begun to bridge this gap \citep{gavrilets2024modelling,gavrilets_coevolution_2021,doi:10.1073/pnas.2504339122,doi:10.1073/pnas.2525139123}, and integrating this into our framework is an important direction for future work. However we note that in the specific context of online communities, focus on the kinds of subjective, psychological forces discussed here is realistic.

In analyzing our model we focus first on the conditions under which extreme, runaway polarization will arise, and second on the capacity of simple interventions to prevent or reverse it. While the structure of our model is complicated, reflecting to some degree the complexity of human identity, belief and decision-making, we also make many simplifications, including the assumption that populations are homogeneous with respect to their utility parameters (although see SI section 2.1), and that opinions are not ``anchored'' with respect to e.g. physical reality \citep{doi:10.1073/pnas.2525139123}.

When simulating interventions, we focus on similar approaches to those that have previously been used to tackle misinformation \citep{Kozyreva24,Coleman22,pennycook_shifting_2021,doi:10.1177/1529100612451018,Lewandowsky03072021}, namely attention nudges and sanctions against accounts that violate platform rules. Among our simulated interventions, attention nudges that amplify moderate elites are the most effective. It is worth noting in this context that a number of authors have highlighted the role of elites who express extreme views in generating polarization in the first place \citep{BETTS2026103179,10.1093/pnasnexus/pgae224,https://doi.org/10.1155/2018/2740959,BACK2023102639}. The realism of our moderate elite intervention depends on the existence of individuals or groups with cross-cutting appeal -- i.e. people who are listened to by ``both sides''. However polarization may undermine the credibility, or even the existence of such elites. 

More generally, the conclusions we draw about interventions must, as with all models, be treated as hypotheses to be explored and tested empirically, not as irrefutable evidence of their efficacy. Nonetheless, our conclusions turn out to be fairly simple. We arrive at a message of hope, and a warning. Our work suggests that polarization can be prevented and even reversed across a wide range of circumstances if influential groups of individuals who possess cross-cutting appeal, and who express moderate attitudes, exist and can be sufficiently amplified. However, reversing expressions of polarization does not necessarily make the attitudes that underpin it disappear. Even where polarization appears to have been successfully reversed, latent extremism may remain.

\section*{Methods}

Here we give details of quantities used by the model and described in the main text, as well as further details on classification of states and interventions.

\subsection*{Memory, utility and distance calculations} As described in the main text, each individual $i$ is characterized by an attitude vector $A^i=(a^i_1,a^i_2,\ldots, a^i_n)$ where $a^i_j$ is $i$'s attitude on topic $j$, and $n$ is the total number of topics under discussion. Each individual is also characterized by two memory vectors. The self-memory vector, $S^i=(s_{j_1},s_{j_2}\ldots s_{j_s})$ contains the $s$ most recent memories of the focal individual's own expressed attitudes, where we denote $s_{j_l}$ to be the self-memory of the focal individual and $j_l$ indicates the attitude dimension the memory relates to. The other-memory vector, $M^i=(m^{k_1}_{j_1},m^{k_2}_{j_2}\ldots m^{k_m}_{j_m})$,  contains the $m$ most recent memories of other individuals' expressed attitudes, where we denote $m^{k_l}_{j_l}$ to be the memory of the focal individual about the attitude of individual $k_l$ on attitude dimension $j_l$. 

The in-group assignment of focal individual $i$ is based on  the average distance $\bar{d}_i$ in attitude space between their own attitude and the attitudes expressed by all other members of the population held in their memory, $M^i$, across all attitude dimensions. i.e.

$$
\bar{d}_i=\frac{1}{m}\sum_{j=1}^n\sum_{l=1}^m (|a^i_j-m_{j_l}^{k_l}|)\delta_{j,j_l}
$$
Where $\delta_{j,j_l}$ is the Kronecker delta, which takes values $\delta_{j,j_l}=1$ when $j=j_l$ and 0 otherwise. We then calculate the average distance of each other individual $k$ from the focal individual, $d_{ik}$, i.e.

$$
d_{ik}=\frac{1}{m_k}\sum_{l=1}^m (|a^i_{j_l}-m_{j_l}^{k_l}|)\delta_{k,k_l}.
$$

Here $m_k$ is the number of times individual $k$ appears in the memory of $i$. Note that we exclude individuals who do not appear in $i$'s memory from the calculation. We then assign all individuals $k$ for who $d_{ik}\leq \bar{d}_i$ to be the in-group of $i$, and all other individuals to the out-group. 

When calculating utility (Eq. 1) the following distance measures are used:
\\
\\
1) The distance $d^j_{\text{in}}$ is the average distance between the focal individual's attitude on topic $j$ and all members $k$ of their in-group $\mathcal{I}$ contained in their memory $M^i$, i.e.

$$
d^j_{\text{in}}=\frac{1}{M^j_{\text{in}}}\sum_{k\in \mathcal{I}}\sum_{l=1}^m (|a^i_{j}-m_{j_l}^{k_l}|)\delta_{j,j_l}
$$
where $M^j_{\text{in}}$ is the number of memories of in-group members' attitudes in $M^i$ on topic $j$. 
\\
\\
2) The distance $d^j_{\text{out}}$ is the average distance between the focal individual's attitude on topic $j$ and all members $k$ of their out-group $\mathcal{O}$ contained in their memory $M^i$, i.e.

$$
d^j_{\text{out}}=\frac{1}{M^j_{\text{out}}}\sum_{k\in \mathcal{O}}\sum_{l=1}^m (|a^i_{j}-m_{j_l}^{k_l}|)\delta_{j,j_l}
$$
where $M^j_{\text{out}}$ is the number of memories of out-group members' attitudes in $M^i$ on topic $j$.
\\
\\
3) The distance $d^j_{\text{self}}$ is the average distance between the focal individual's current attitude on topic $j$ and all of their self-memories $S^i$, i.e.

$$
d^j_{\text{self}}=\frac{1}{M^j_{\text{self}}}\sum_{l=1}^s (|a^i_{j}-s_{k_l}|)\delta_{j,k_l}
$$

where $M^j_{\text{self}}$ is the number of self-memories in $S^i$ on topic $j$.

\subsection*{Hard and soft Overton window}

The hard and soft Overton windows are both defined as ranges in attitude space outside of which individuals incur a utility penalty if they express an attitude. The width of the hard Overton window $w_{H}$ is fixed for the duration of simulations (unless it is changed as part of an intervention). Note that we always assume that the hard Overton window is centered on 0. The penalty associated with the hard Overton window is $c_H$, and this is applied to the utility calculation for an attitude $|a^i_j|>w_H/2$

The soft Overton window functions in the same way except that it is based on the range of attitudes held by all members of the population on topic $j$. The penalty associated with the soft Overton window is $c_S$, and this is applied to the utility calculation for an attitude $a^i_j>a^{\text{max}}_j$ or $a^i_j<a^{\text{min}}_j$ where $a^{\text{max}}_j$ is the maximum value of an expressed attitude on topic $j$ in the memory of the focal individual and $a^{\text{min}}_j$ is the minimum value.

\subsection*{Classification of states}
We classify the population level outcome state of simulations as either consensus, constrained polarization, or runaway polarization. In order to do this we first classify each of a focal individual $i$'s $D$ attitudes associated with each identity dimension on a given topic $j$ to be either ``active'' or ``inactive''. A given dimension is classified as active on topic $j$ if the associated attitude was expressed at least 2\% of the times that $i$ expressed an attitude on $j$, and is otherwise classified as inactive. This calculation is performed over the most recent $200,000$ steps of the model (i.e. 400 days in our time units). 

We define a population to be in the runaway polarization state if 5\% of ``active'' attitudes 
are greater than 1 unit outside the hard Overton window, i.e. $|a^i_j|>w_{H}/2+1$. 

We define a population to be in a state of constrained polarization using the following rule: 
\\
\\
i) For each attitude dimension, find the attitudes corresponding to the most used identity dimension of each individual (which we call the dominant identity). 
\\
\\
ii) Calculate the average absolute deviation from the mean across all individuals for that attitude dimension, 

$$
\rho^j=\frac{1}{N}\sum_{i=1}^N|\bar{a}^j-a^i_j|
$$ 

where $\bar{a}^j=\frac{1}{N}\sum_{i=1}^Na^i_j$.
\\
\\
iii) Separately calculate the average absolute deviation from the mean for attitudes larger and smaller than the average attitude 

$$
\rho^j_{<}=\frac{1}{N_{<}}\sum_{i=1}^{N_{<}}|\bar{a}^j_{<}-a^i_j|
$$ 

$$
\rho^j_{>}=\frac{1}{N_{>}}\sum_{i=1}^{N_{>}}|\bar{a}^j_{>}-a^i_j|
$$ 

Where $N_{>}$ and $N_{<}$ are the number of individuals whose attitude associated with their dominant identity lies above or below $\bar{a}_j$ respectively. Similarly $\bar{a}^j_>$ and $\bar{a}^j_<$ are the average attitudes of those whose attitude along the dominant identity dimensions lie above or below $\bar{a}_j$.
\\
\\
iv) Calculate the ratios $\frac{\rho^j_{<}}{\rho^j}$ and $\frac{\rho^j_{>}}{\rho^j}$.
\\
\\
v) The smaller the ratios are the less likely it is that attitudes are unimodally distributed, we choose a threshold of $0.3$ which is low enough that $100$ draws from a uniform distribution will be misclassified as polarized with an error rate of 1 in $10^8$ simulations. 
\\
\\
Any population that does not meet this criterion is defined to be in a state of consensus. Applying these two classification rules (for escaping the hard Overton window and being in a state of polarization) yields the three classifications of states given in Figure 2. We do not observe ``runaway consensus'' for the parameters considered, although this could of course occur if the cost for violating the hard Overton window, $c_H$, is made sufficiently small

\subsection*{Interventions}
\begin{enumerate}
    \item The cost hike intervention is implemented as an increase of the cost of transgressing the hard Overton window $c_H^{int} = c_H^{pre} + c_{hike}$. In our simulated interventions $c_H^{pre} = 0.5$ and $c_H^{hike} = 50$, corresponding to a 100 fold increase. We observe that, for a wide range of parameters, a cost hike of $1<c_H^{hike} < 10$ is sufficient to prevent runaway polarization (see Figure S20). 
    \item The hard Overton window optimization intervention works by extending or shrinking the width of the hard Overton window at initialization, $w_H^{pre}$, such that the hard Overton window width at the time of the intervention is $w_H^{int}=3.0$. This puts $w_H^{int}$ at the point where the consensus outcome state is most prevalent (see Figure 3).
    \item The attention balance intervention sets attention bias from $\alpha^{pre}$ at initialization to $\alpha^{int}=0$ at time of intervention. This ensures that after the intervention, individuals are equally likely to begin new threads on both topics.
    \item The amplifying moderate elites intervention introduces an additional term to the utility function (Eq. 1) corresponding to a pull towards the center of the hard Overton window. This corresponds to a group  of influential individuals, who are perceived by all individuals as part of their in-group, and who express moderate attitudes. In particular, we introduce a term $\lambda_{\text{elite}} d_{\text{elite}}^j$ which is subtracted from Eq. 1, where $d_{\text{elite}}=|a^i_j|$. The coefficient $\lambda_{\text{elite}}$ before intervention is set to $\lambda_{\text{elite}}^{pre}=0$ and post intervention to $\lambda_{\text{elite}}^{int}=1$. We observe that, for a wide range of parameters, a value of $\lambda_{\text{elite}}^{int}>0.3$ is sufficient to prevent runaway polarization (see Figure S19). 
\end{enumerate}

\subsection*{Latent extremism}

When implementing the cost hike and moderate elites interventions, we noticed that in response individuals in our model would express those of their attitudes that lie within the Overton window in favor of those that transgress it. That meant that while runaway polarization was effectively reversed, the system retained ``latent extremeness''. To track the extent of such latent extremeness we define the measure 

\[
e_j = \frac{1}{D}\sum^{D}_{i=1}\left(|a_j^i| -\frac{w_{H}}{2}\right)\left(1-f^i_j\right)
\] 

where $f^i_j$ is the proportion of times over the past $200,000$ time steps (i.e. 400 days in our time units) that identity dimension $i$ was used for individual $j$. 
The measure is thus a weighted average of how far out side of the Overton window attitudes lie when they are not expressed.

\newpage

\section*{Supplementary Information}

\tableofcontents


\newpage

\section{Additional model details}

Here we give additional details on the parameters used and the decisions made in constructing the model presented in the main text.

\subsection{Parameter choices}

As described in the main text our model comprises a range of different types of parameters which can be roughly grouped into three categories: i) Utility parameters. ii) Cognitive parameters. iii) Platform parameters. Utility parameters ($\lambda_{\text{self}}$, $\lambda_{\text{pull}}$, $c_S$, and $r$) describe the weights, endogenous costs and update step-size associated with the utility function (i.e. the preferences of the individual being modeled). Cognitive parameters ($m$, $n$, $D$, and $s$) describe the cognitive capacity of an individual (i.e. the memory capacity and dimensionality of attitudes and identities retained by the individual being modeled). Platform parameters ($c_H$, $\beta$ and $w_H$) describe the parameters that we conceive as being exogenously set by the social media platform being modeled. 

Clearly these categories are only rough descriptions, and there is overlap between the three (e.g. $c_H$ is both a utility parameter and something we assume is endogenously set by the platform). In the main text and in the SI below, we treat variations in the platform parameters as corresponding to interventions, and we assess the impact of those interventions across a wide range of choices for the utility and cognitive parameters. We do this first by varying pairs of parameters keeping others fixed (as in main text Figure 3). When we do this we set the other parameters to ``default'' values. These default values are given in Table S1 below. They are chosen to be ``intermediate'', meaning that, for these choices, different dynamics may be easily observed by varying one or two of the parameters, keeping the other fixed. The choices are based on computational exploration of the model dynamics.

\begin{center}
\captionof*{table}{\textbf{Table S1}. Default parameters for the model}
\begin{tabular}{r|p{8cm}|r}
  \textbf{Parameter} & \textbf{Description} & \textbf{Default value} \\
  \hline
  $N$ & The number of individuals in the simulation & 100 \\
  $n$ & The number of attitude dimensions & 2 \\
  $D$ & The number of identity dimensions & 2 \\
  $m$ & The number of memories of others & 50 \\
  $s$ & The number of self-memories & 25 \\
  $r$ & The size of attitude updates as a proportion of distance to the expressed attitude & 0.2 \\
  $\lambda_{\text{self}}$ & The self-weight parameter in the utility function & 0.4 \\
  $\lambda_{\text{pull}}$ & The pull parameter in the utility function & 0.4 \\
  $c_S$ & The cost of transgressing the soft Overton window & 0.5 \\
  $c_H$ & The cost of transgressing the hard Overton window & 0.5 \\
  $\alpha$ & $\alpha=2\beta-1$ where $\beta$ is the probability that the dominant attitude dimension used to sample which attitude dimension is used in new threads & 0.5 \\
  $w_H$ & The width of the hard Overton window (for each attitude dimension) & 3.0 \\
\end{tabular}
\end{center}

\newpage

\subsection{Modeling thread length}

To model thread length we assume that, for a given post at given depth $r$, each member of the population of $N$ individuals replies with probability 

$$
p(r)=\frac{1}{N+r}
$$

where we set $r=0$ for the original post. This ensures that a given initial post is expected to get one reply, and the probability of further replies decreases with thread depth. Note that we do not attempt to model heterogeneity in poster ``influence'', which can result in ``mega threads'' and lead to a heavily right-skewed distribution of thread length.

Our model of threads corresponds to a branching process with time dependent birth. In particular, the ``birth rate'' decreases monotonically with depth (i.e. time). When $r=0$ the expected number of replies is 1. If this rate were fixed this would correspond to a critical branching process. At every step after this the birth rate is less than 1. Thus starting at any depth $r>0$, assuming we have $n>0$ threads at that depth, we have $n$ independent sub-critical branching process, and extinction is assured.

Figure S1 shows the distribution of thread lengths resulting from our model under the default parameters. We see that the median thread length is $\sim 3$, consistent with social media data.

\begin{figure}[!h]
    \centering
    \includegraphics[width=0.5\linewidth]{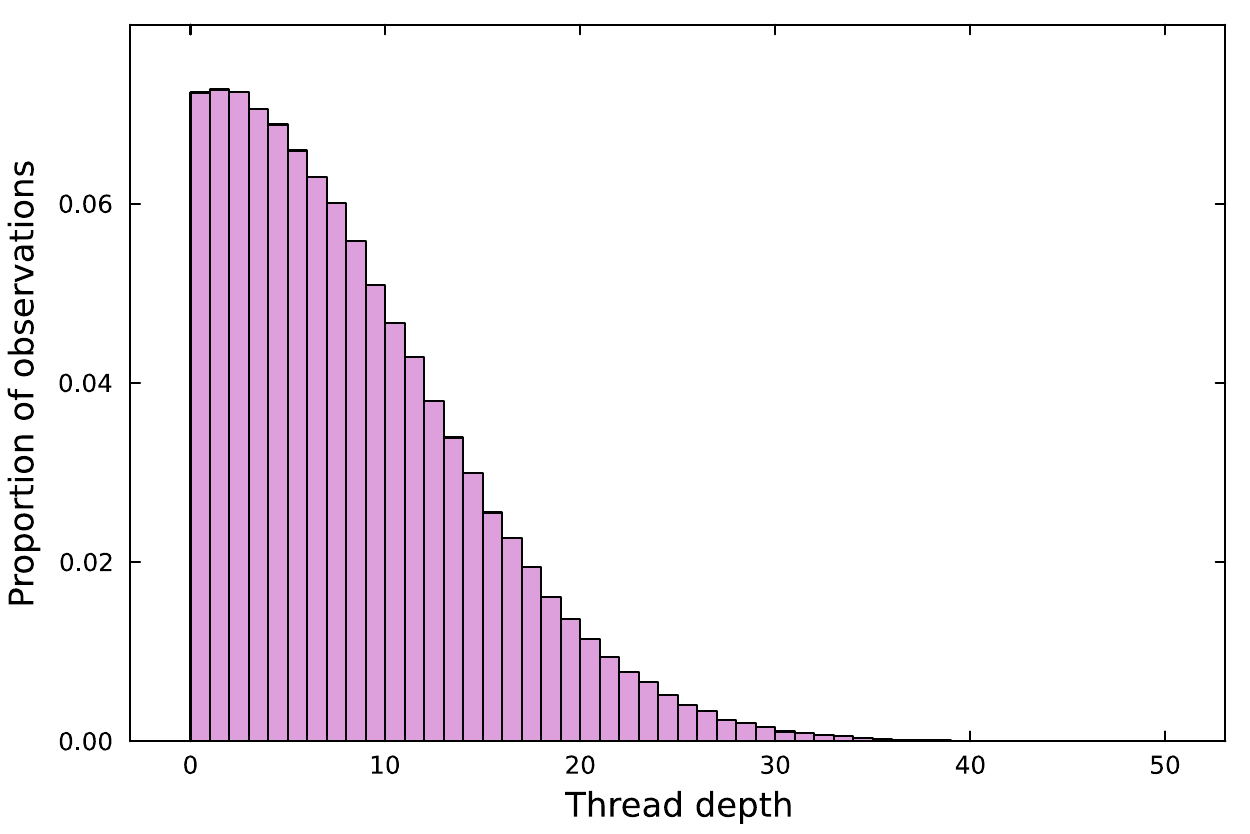}
    \caption*{\textbf{Figure S1}. The distribution of thread depth using our thread generation formula for $100,000$ generated threads.}
    \label{fig:placeholder}
\end{figure}

\newpage

\subsection{Classification of states}

As described in the main text, we classify the dynamics of the model in terms of three outcome states: consensus, constrained polarization and runaway polarization. Here we provide additional robustness checks for the efficacy of our classification method.

\subsubsection{Cluster size}

We calculate the size of the largest cluster for states classified as polarized. We do this for first for states classified as runaway (Figure S2a). We see that in the first case the largest cluster rarely exceeds half the population and never comprises the whole population. Thus we observe that whenever a population breaks free of the hard Overton window, this coincides with the fracturing of the population into two groups (i.e. polarization). Figure S2b shows the same analysis for all states classified as polarized. Here we see that the largest cluster rarely exceeds 80\% of the population and never comprises the whole population. However we do observe cases in which the largest cluster comprises less than half the population, corresponding to the fracturing of the population into $>2$ clusters.

\begin{figure}[!h]
    \centering
    \includegraphics[width=0.8\linewidth]{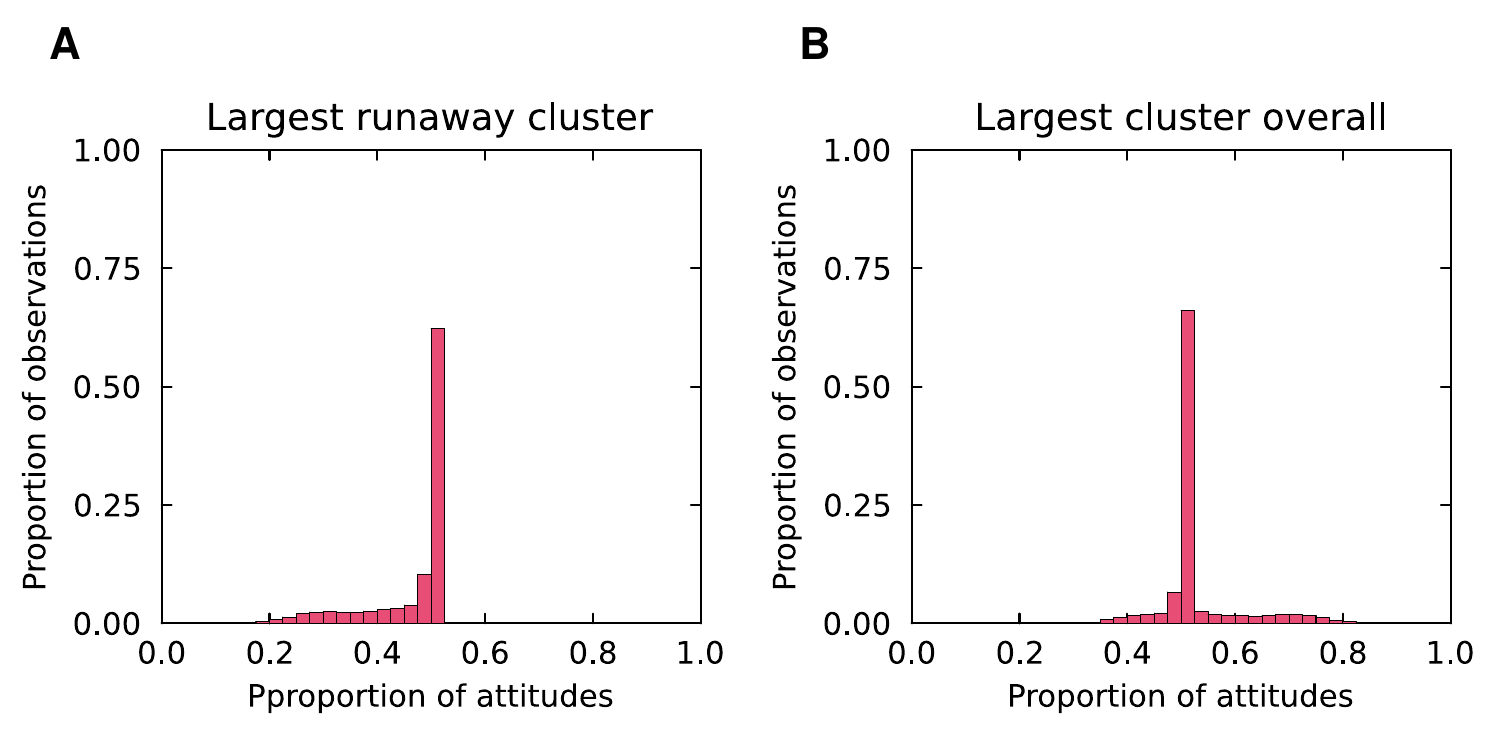}
    \caption*{\textbf{Figure S2}. A histogram of the proportion of attitudes that make up the largest cluster under runaway polarization conditions varying attention bias $\alpha$ and the widht of the hard Overton window $w_H$. Cluster sizes are rarely larger than half of the attitudes in the population and are never observed as close to 100\%, indicating that the model does not produce "runaway consensus". Observations are taken from simulations varying the hard Overton window width and attention bias. Simulations were run for the equivalent of 50 years (10 million attitude updates) with $N=100$ individuals, $n=2$ attitude dimensions, $D=2$ identity dimensions, $m=50$ memories of others, and $s=25$ self-memories. Utility parameters consist of self weight $\lambda_{\text{self}}=0.4$, in-group pull of $\lambda_{\text{pull}}=0.4$, soft Overton cost of $c_S=0.5$, hard Overton cost of $c_H=0.5$, and attitude update step size of $r=0.2$.}
    \label{fig:placeholder}
\end{figure}

\newpage

\subsubsection{Polarization threshold}

As we describe in the Main Text (see Materials and Methods) we calculate a measure of bimodality and assign the outcome state of a simulation to be ``polarized'' if this measure is greater than a threshold value of 0.3, which results in misclassification of a uniformly distributed set of opinions as polarized with a probability of $10^{-8}.$

Below in Figure S3, we calculate the same error rate as a function of the threshold and as a function of the number of opinions being sampled (which corresponds to the number of interlocutors in our simulations). We see that error rates are robust for threshold parameters $<0.4$ and for populations $>2^5$ (i.e. $>30$ od so individuals).

\begin{figure}[!h]
    \centering
    \includegraphics[width=0.8\linewidth]{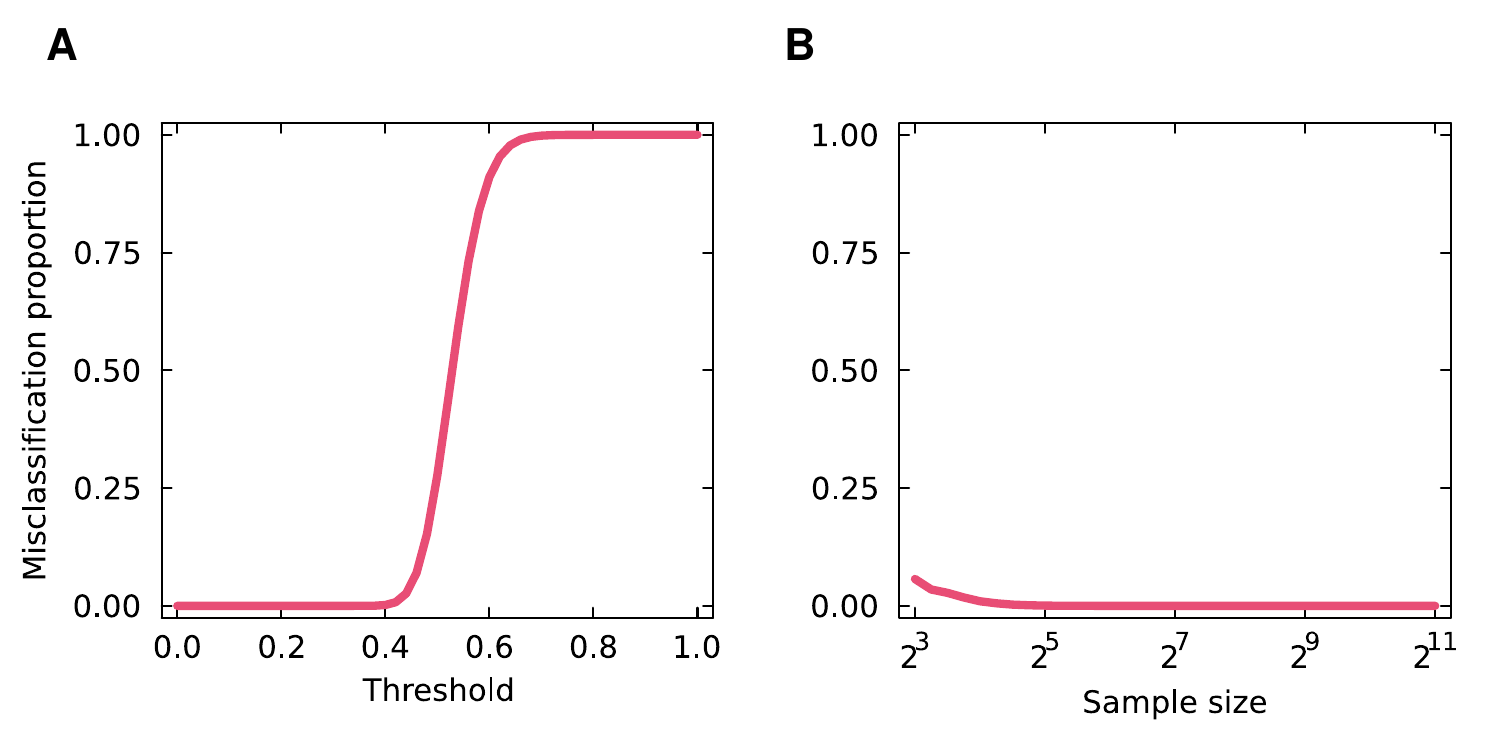}
    \caption*{\textbf{Figure S3}. The proportion of draws from a uniform distribution that result in a false positive constrained polarization classification using our classification method varying the threshold parameter (panel A) and the number of draws from the uniform distribution (panel B). Panel A shows that choosing a threshold value of $0.3$ for a population size of $N=100$ makes classification very unlikely to fail with zero misclassifications over ten million replicates. Panel B further shows that a threshold value of $0.3$ performs well also for smaller population sizes.}
    \label{fig:placeholder}
\end{figure}

\newpage

\section{Additional simulation results}

Here we present additional results from simulations. We systematically vary utility parameters, cognitive parameters and interventions, in order to assess the robustness of the results presented in the main text.

\subsection{Utility parameters}

Utility parameters comprise those that are part of the utility function (Main Text Eq. 1), and therefore capture the ``goals'' of the individual being modeled. 

\subsubsection{Self-consistency \& in-group pull}

In the main text we vary the two key weight parameters $\lambda_{\text{self}}$ and $\lambda_{\text{self}}$ which capture, respectively, the strength of desire for self consistency in expression vs group conformity, and the strength of desire for in-group similarity vs out-group differentiation. Here we capitulate main-text Figure 3a for different values of hard Overton window width, $w_H$ and attention bias $\beta$. We see that outcome states in these cases are similar, with the potential for constrained polarization being somewhat sensitive to these parameters. 

\begin{figure}[!h]
    \centering
    \includegraphics[width=0.8\linewidth]{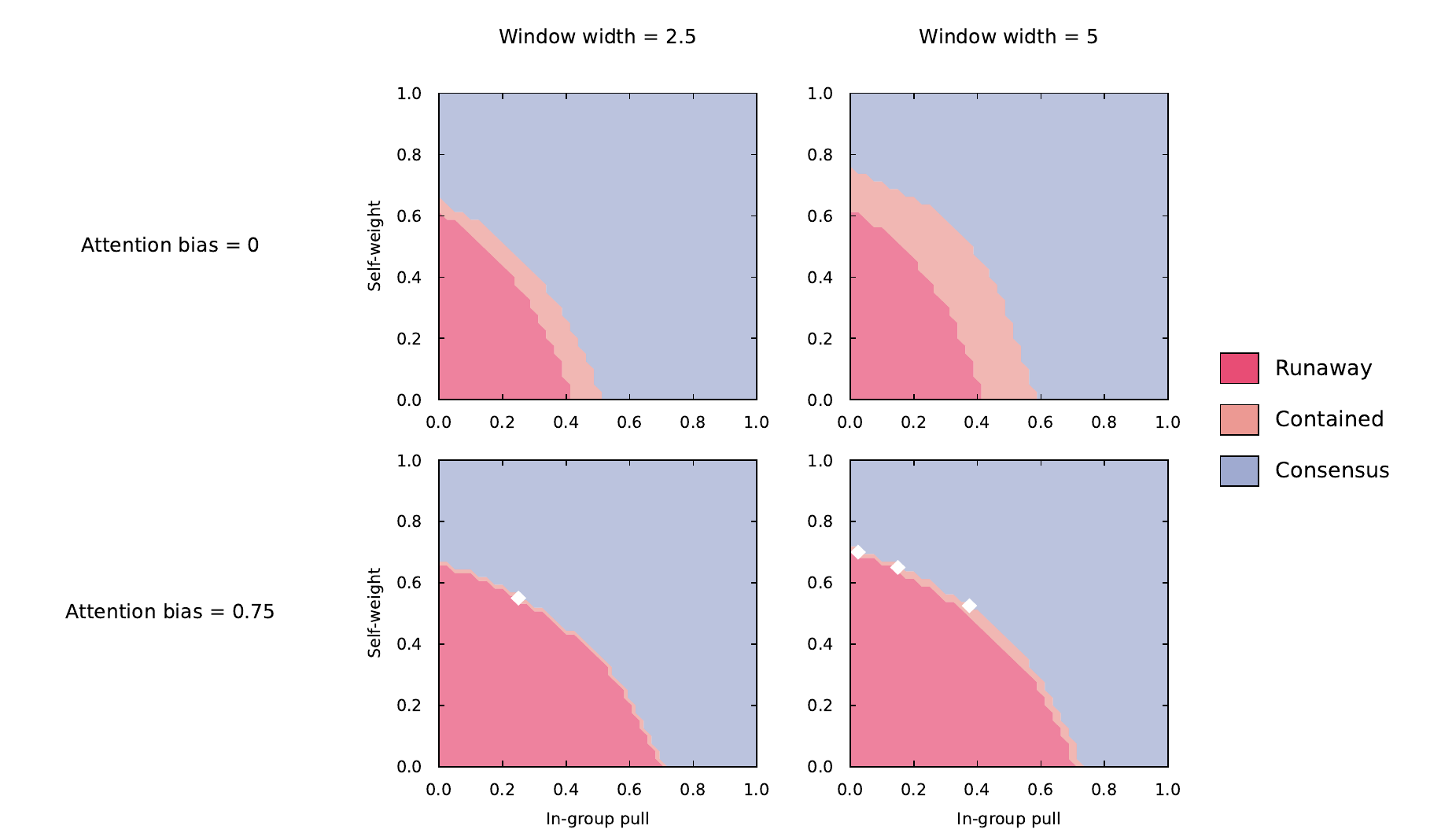}
    \caption*{\textbf{Figure S4}. The dominant outcome state across values of in-group pull and self-weight utility parameters for high and low values of attention bias $\alpha$ and the hard Overton window width $w_H$. Simulations were run for the equivalent of 50 years (10 million attitude updates) with $N=100$ individuals, $n=2$ attitude dimensions, $D=2$ identity dimensions, $m=50$ memories of others, $s=25$ self-memories and attitude update step size of $r=0.2$. Hard and soft Overton costs are $c_H=0.5$ and $c_S=0.5$. A given set of parameters is classified as a particular state if they result in that classification at least 50\% of the time over 200 replicate simulations with initial conditions randomized withing the hard Overton window. White regions corresponds to cases where no outcome state meets this threshold.}
\end{figure}

\newpage

\subsubsection{Update step size}

Next we vary the update step size, $r$. While this is not an input to the utility function, it determines how much an individual's attitude, and therefore their utility, changes with each update. We see (Figure S5), that as $r$ increases, runaway polarization becomes more likely.

\begin{figure}[!h]
    \centering
    \includegraphics[width=\linewidth]{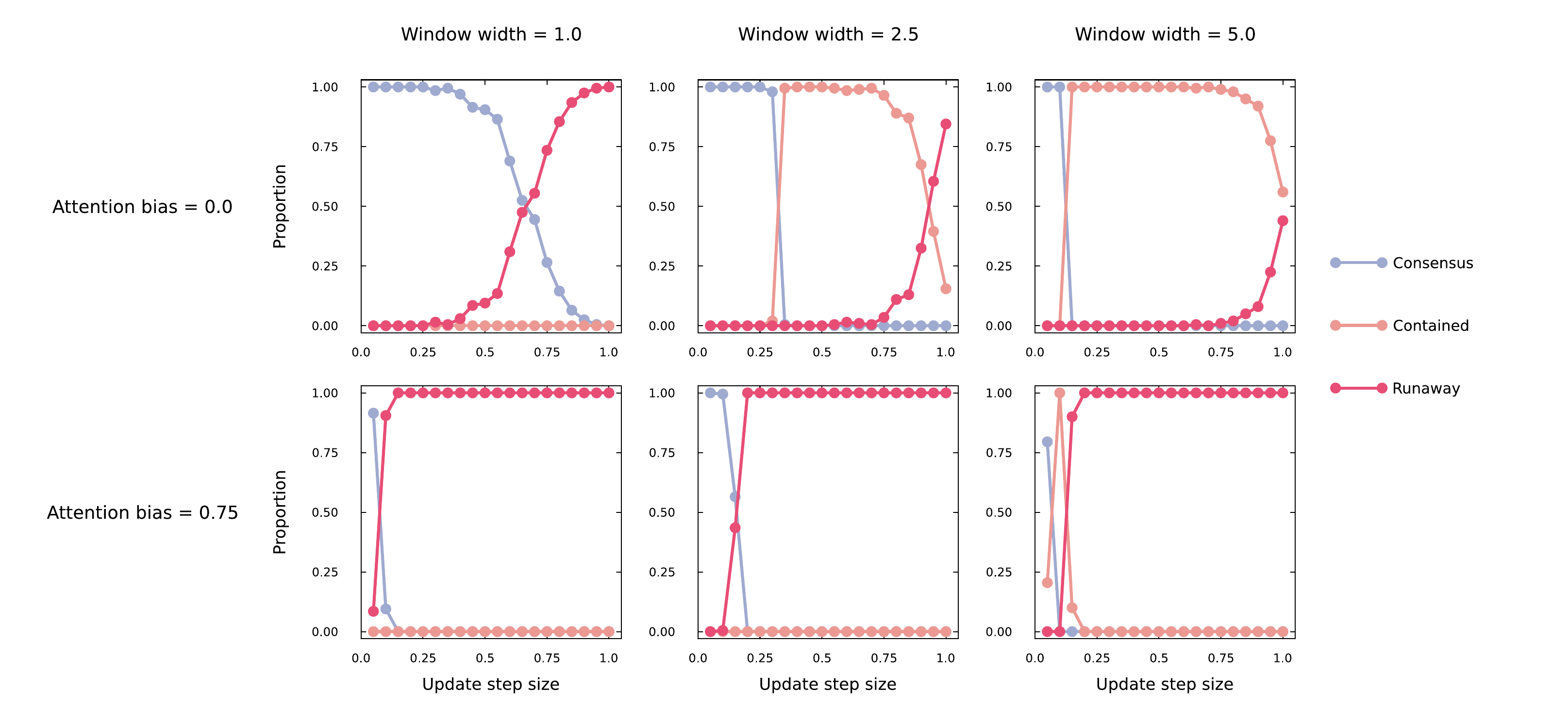}
    \caption*{\textbf{Figure S5}. The proportion of consensus, constrained polarization, and runaway polarization, for high and low values of Overton window width and Attention bias, varying the relative step size for updating attitudes in the simulation $r$. Simulations were run for the equivalent of 50 years (10 million attitude updates) with $N=100$ individuals, $n=2$ attitude dimensions, $D=2$ identity dimensions, $m=50$ memories of others, and $s=25$ self-memories. Utility parameters consist of self weight  $\lambda_{\text{self}}=0.4$, an in-group pull of $\lambda_{\text{pull}}=0.4$, soft Overton cost of $c_S=0.5$ and hard Overton cost of $c_H=0.5$.}
\end{figure}

\newpage

\subsubsection{Population heterogeneity}

Next we relax the assumption that utility parameters, $\lambda_{\text{self}}$ and $\lambda_{\text{self}}$, are the same for all interlocutors in a given simulation. We drew these parameters instead from a normal distribution and systematically increased the standard deviation, for fixed mean. Figure S6 shows the result of increasing standard deviation in these parameters. We see that once it reaches a threshold of $\sigma\sim 0.05$ is reached runaway polarization tends to replace constrained polarization or consensus. This suggests that polarization ``spreads'', with those individuals who are more susceptible (i.e. with low $\lambda_{\text{self}}$ and $\lambda_{\text{self}}$) becoming polarized and impacting the rest of the, less-susceptible, population. This is similar to the dynamics found in \citep{doi:10.1073/pnas.2102140118,doi:10.1126/sciadv.abd4201}.

\begin{figure}[!h]
    \centering
    \includegraphics[width=\linewidth]{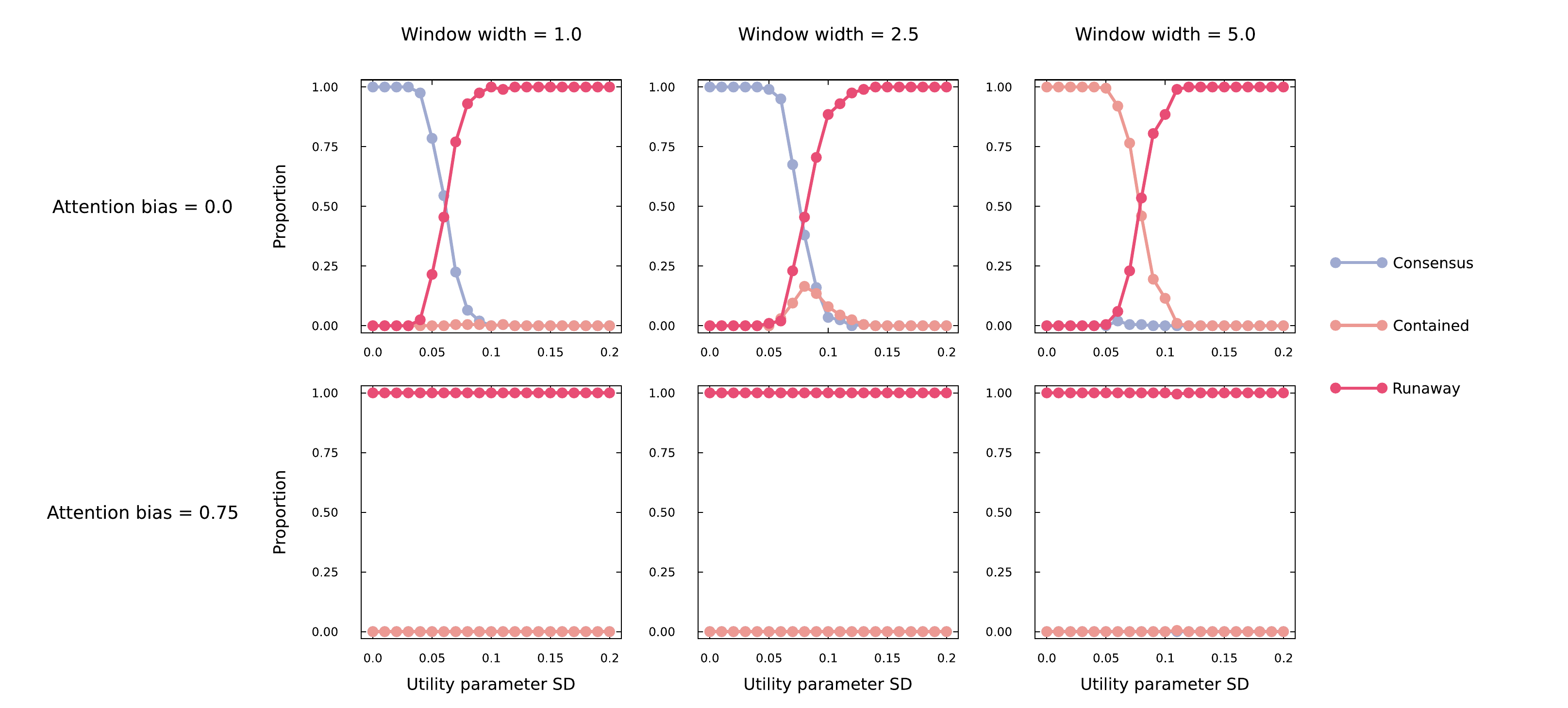}
    \caption*{\textbf{Figure S6}. The proportion of consensus, constrained polarization, and runaway polarization, for high and low values of Overton window width and Attention bias. Instead of letting all individuals have the same utility parameter values, this figure shows outcome states varying the standard deviation $\sigma$ of a Gaussian used to sample parameter values for each individual. The in-group pull $\lambda_{\text{pull}}$ and the self-weight $\lambda_{\text{self}}$ parameters were sampled from $\mathcal{N}(\lambda, \sigma^2)$. Simulations were run for the equivalent of 50 years (10 million attitude updates) with $N=100$ individuals, $n=2$ attitude dimensions, $D=2$ identity dimensions, $m=50$ memories of others, and $s=25$ self-memories. Utility parameters consist of self weight $\lambda_{\text{self}}=0.4$, in-group pull of $\lambda_{\text{pull}}=0.4$, soft Overton cost of $c_S=0.5$, hard Overton cost of $c_H=0.5$, and attitude update step size of $r=0.2$.}
\end{figure}

\newpage

Next we systematically varied the standard deviation in the cost parameters, $c_S$ and $c_H$, which corresponds to variation in the levels of susceptibility to sanction (i.e. how much a given individual cares about being told off or having their post deleted etc). Figure S7 shows the results, which are qualitatively similar to Figure S6, with a threshold standard deviation $\sigma\sim 0.1$. As in the previous scenario, this suggests that polarization spreads, with individuals less susceptible to sanction initially becoming polarized, then impacting the rest. 

\begin{figure}[!h]
    \centering
    \includegraphics[width=\linewidth]{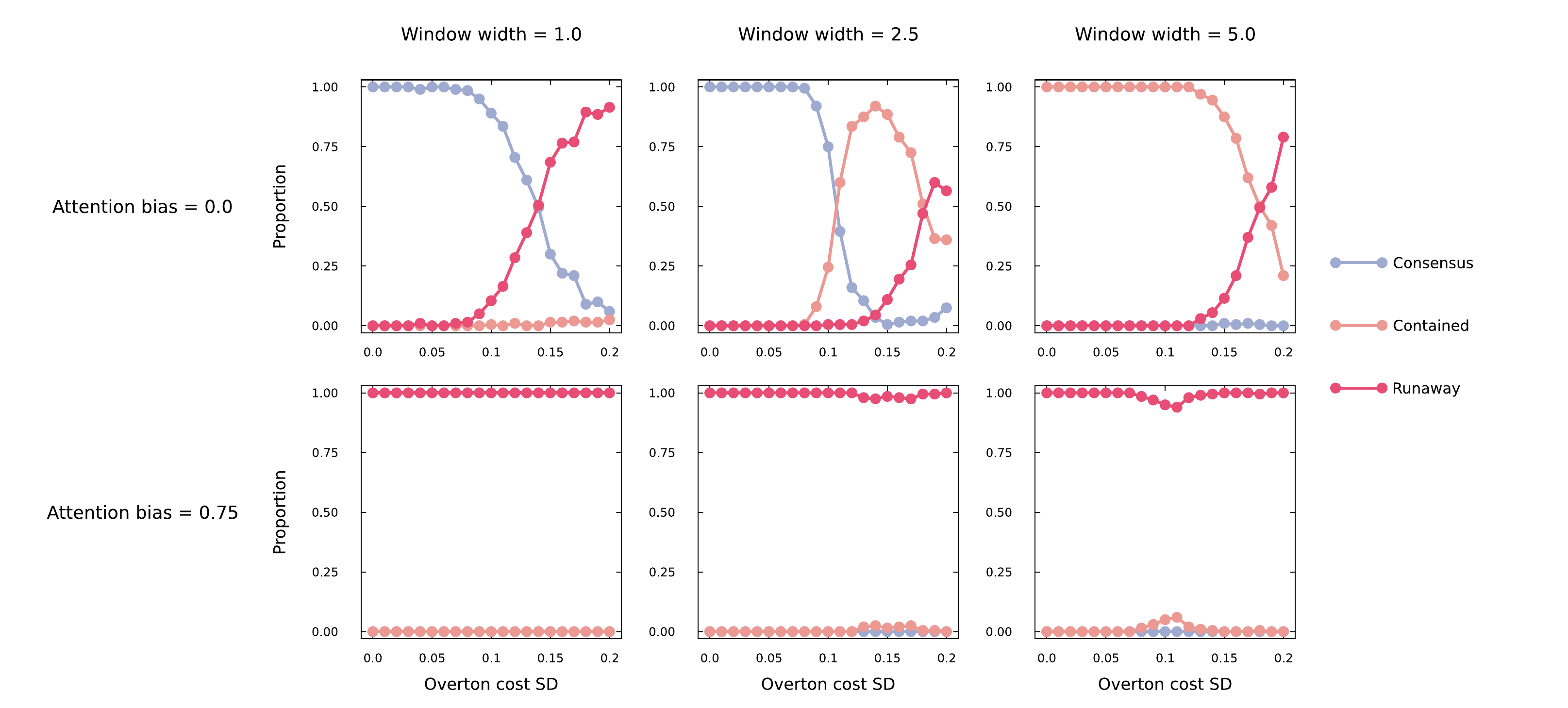}
    \caption*{\textbf{Figure S7}. The proportion of consensus, constrained polarization, and runaway polarization, for high and low values of Overton window width and Attention bias. Instead of letting all individuals have the same utility parameter values, this figure shows outcome states varying the standard deviation $\sigma$ of a Gaussian used to sample parameter values for each individual. The hard and soft Overton window cost $c_H,\,c_S$ parameters were sampled from $\mathcal{N}(c, \sigma^2)$. Simulations were run for the equivalent of 50 years (10 million attitude updates) with $N=100$ individuals, $n=2$ attitude dimensions, $D=2$ identity dimensions, $m=50$ memories of others, and $s=25$ self-memories. Utility parameters consist of self weight $\lambda_{\text{self}}=0.4$, in-group pull of $\lambda_{\text{pull}}=0.4$, soft Overton cost of $c_S=0.5$, hard Overton cost of $c_H=0.5$, and attitude update step size of $r=0.2$.}
 \end{figure}

\newpage

\subsubsection{Combinations of parameters}

We explored combinations of seven parameters: Attention bias, $\beta$, hard Overton window size, $w_H$, in-group pull of $\lambda_{\text{pull}}$, self weight $\lambda_{\text{self}}$, update step-size $r$, soft Overton window cost $c_S$, and hard Overton window cost $c_H$.

We set these chosen to correspond to ``HIGH'', ``MEDIUM'' or ``LOW'' values, resulting in 1458 different ``experimental conditions''. For each experiment we measure the proportion of 200 replicate simulations that result in runaway polarization. The results are shown in Table S1, which is attached as a spreadsheet due to its size. While it is not possible to briefly summarize the interactions between all parameters, this serves as a lookup table for those interested in specific, qualitative scenarios.

\newpage

\subsection{Cognitive parameters}

Next we explore variation in the cognitive parameters, i.e. those that impact the type and amount of information retained by a given individual.

\subsubsection{Number of attitude dimensions}

First we explore variation in the number of attitude dimensions, i.e. the number of different topics an individual can choose from when starting a discussion, and on which they retain an opinion. In order to model attention bias with $>2$ attitude dimensions, we assume that there is a single ``dominant'' attitude dimensions and that, when a non-dominant topic is selected to start a thread, we choose from all non-dominant attitude dimensions equally. Figure S8 shows the impact of varying the number of attitude dimensions. We see that as the number increases, runaway polarization tends to become more likely. This is likely because each of the non-dominant attitudes become even less frequently used as the number of attitude dimensions increases, which is analogous to increasing attention bias with two attitude dimensions (i.e. Main Text Figure 3). 

\begin{figure}[!h]
    \centering
    \includegraphics[width=\linewidth]{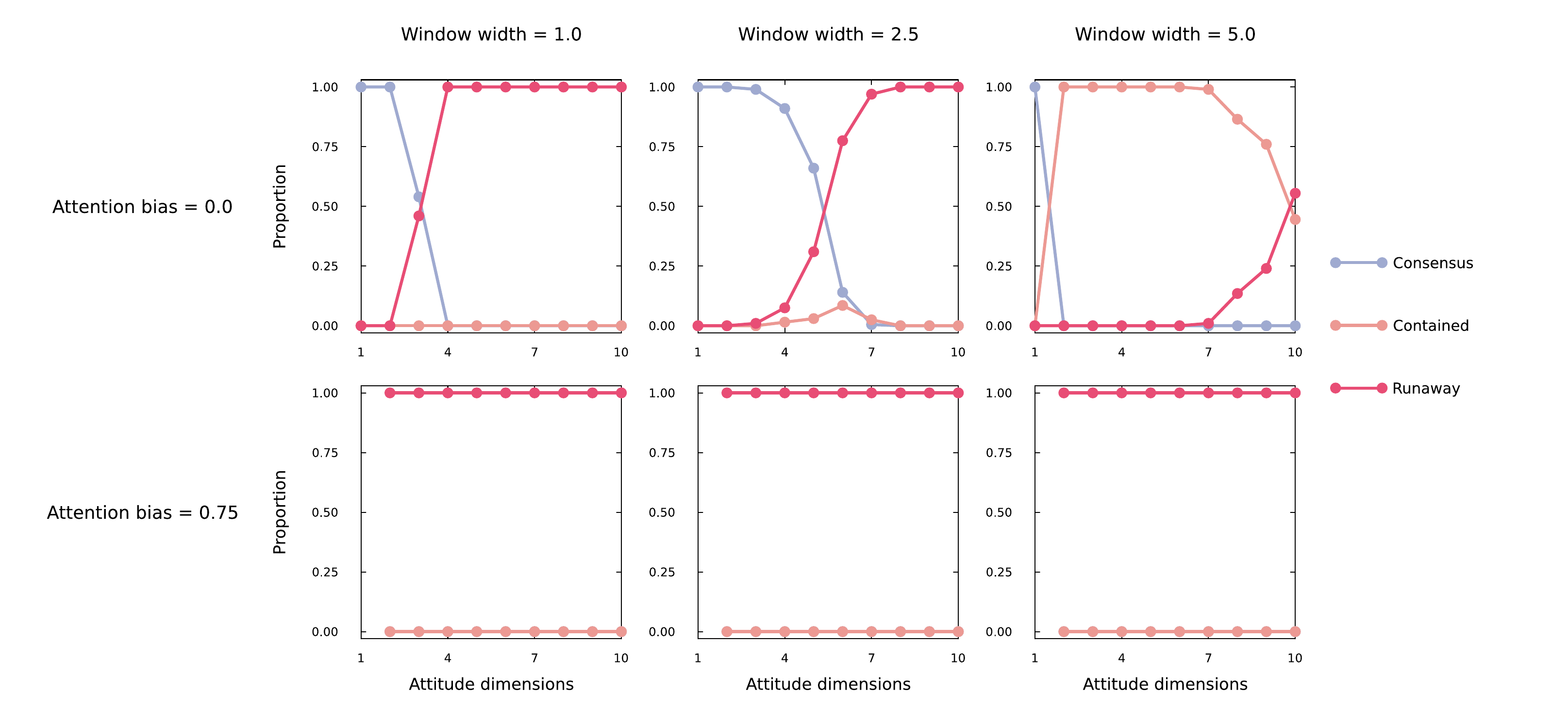}
    \caption*{\textbf{Figure S8}. The proportion of consensus, constrained polarization, and runaway polarization, for high and low values of Overton window width and Attention bias, varying the number of attitude dimensions $n$ in the simulation. Simulations were run for the equivalent of 50 years (10 million attitude updates) with $N=100$ individuals, $D=2$ identity dimensions, $m=50$ memories of others, $s=25$ self-memories and relative attitude update step size of $r=0.2$. Utility parameters consist of self weight  $\lambda_{\text{self}}=0.4$, an in-group pull of $\lambda_{\text{pull}}=0.4$, soft Overton cost of $c_S=0.5$ and hard Overton cost of $c_H=0.5$.}
\end{figure}

\newpage

\subsubsection{Number of identity dimensions}

We then further varied the number of identity dimensions, $D$. In Particular we remade main text Figure 3b for $D=1$ and $D=10$ identity dimensions respectively (Figure S9). We see that the qualitative features of the figure are largely insensitive to increasing or decreasing the number of identity dimensions.

\begin{figure}[!h]
    \centering
    \includegraphics[width=0.8\linewidth]{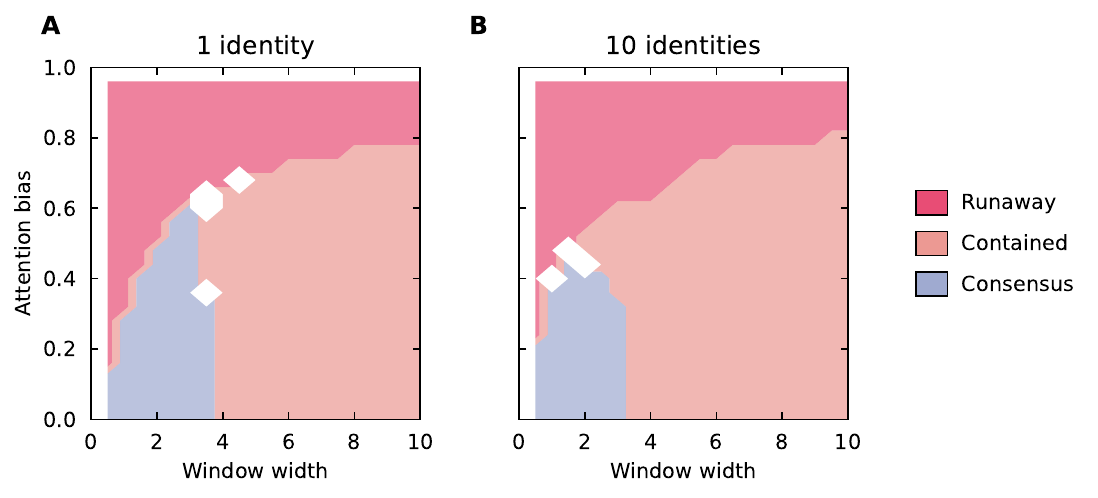}
    \caption*{\textbf{Figure S9}. The dominant outcome state across values of attention bias and hard Overton window width for (A) 1 identity dimension and (B) 10 identity dimensions. Simulations were run for the equivalent of 50 years (10 million attitude updates) with $N=100$ individuals, $n=2$ attitude dimensions,  $m=50$ memories of others, $s=25$ self-memories and attitude update step size of $r=0.2$. Utility parameters consist of self weight $\lambda_{\text{self}}=0.4$, in-group pull $\lambda_{\text{pull}}=0.4$; soft and hard Overton costs are set to $c_S=0.5$ and $c_S=0.5$. A given set of parameters is classified as a particular state if they result in that classification at least 50\% of the time over 200 replicate simulations with initial conditions randomized withing the hard Overton window. White regions corresponds to cases where no outcome state meets this threshold or where there is no data.}
\end{figure}

\newpage

We find similar results when we systematically increase the number of identity dimensions from 1 to 10 (Figure S10). Note that while the number of identity dimensions does not appear to impact the outcome state of the model for fixed parameters, it does impact the efficacy of some interventions (see Main Text Figure 5).

\begin{figure}[!h]
    \centering
    \includegraphics[width=\linewidth]{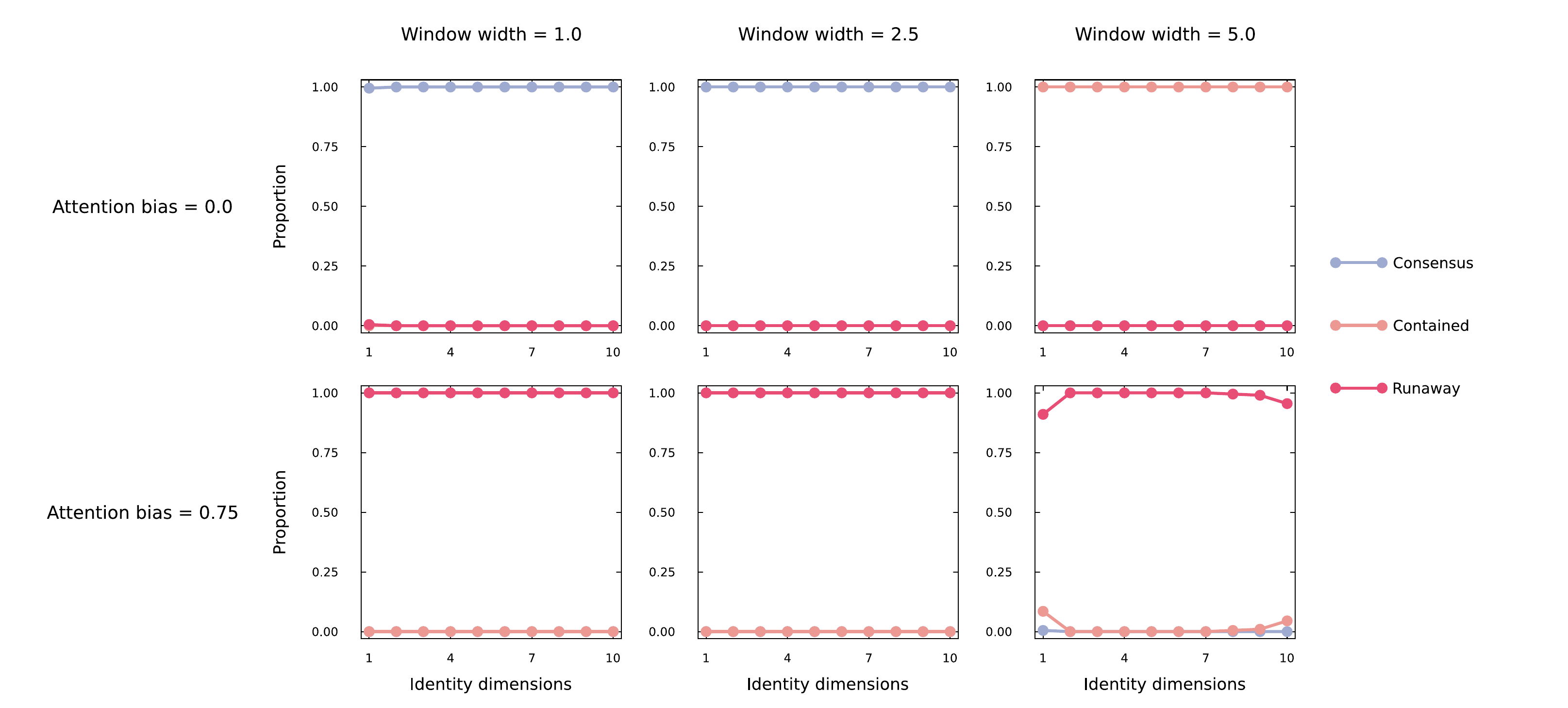}
    \caption*{\textbf{Figure S10}. The proportion of consensus, constrained polarization, and runaway polarization, for high and low values of Overton window width and Attention bias, varying the number of identity dimensions $D$ in the simulation. Simulations were run for the equivalent of 50 years (10 million attitude updates) with $N=100$ individuals, $n=2$ attitude dimensions, $m=50$ memories of others, $s=25$ self-memories and relative attitude update step size of $r=0.2$. Utility parameters consist of self weight $\lambda_{\text{self}}=0.4$, an in-group pull of $\lambda_{\text{pull}}=0.4$, soft Overton cost of $c_S=0.5$ and hard Overton cost of $c_H=0.5$.}
\end{figure}

\newpage

\subsubsection{Number of interlocutors}

While not strictly a ``cognitive'' parameter, we also include the impact of varying the number of interlocutors, $N$, in this section, as this impacts the amount of memory available for a given individual (which can be seen as a cognitive parameter). Figure S11 shows that, increasing the number of interlocutors in a given conversation, tends to make runaway polarization more likely.

\begin{figure}[!h]
   \centering
   \includegraphics[width=\linewidth]{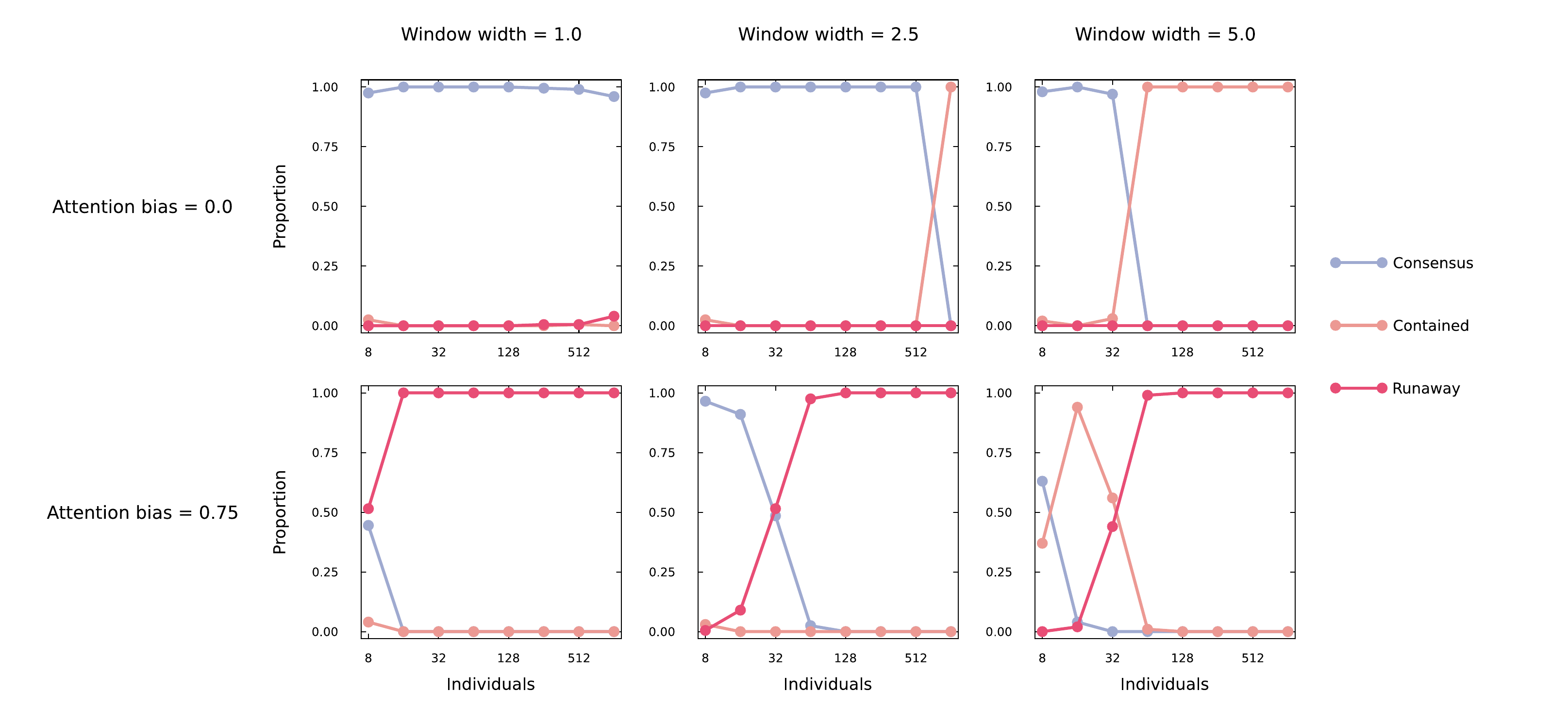}
   \caption*{\textbf{Figure S11}. The proportion of consensus, constrained polarization, and runaway polarization, for high and low values of Overton window width and Attention bias, varying the number of individuals $N$ in the simulation. Simulations were run for the equivalent of 50 years (10 million attitude updates) with $n=2$ attitude dimensions, $D=2$ identity dimensions, $m=50$ memories of others, $s=25$ self-memories and relative attitude update step size of $r=0.2$. Utility parameters consist of self weight $\lambda_{\text{self}}=0.4$, an in-group pull of $\lambda_{\text{pull}}=0.4$, soft Overton cost of $c_S=0.5$ and hard Overton cost of $c_H=0.5$.}
\end{figure}

\newpage

\subsubsection{Memory length}

We systematically varied the number of self memories, $s$ retained by each individual. As shown in Figure S12, longer self-memory tends to result in less runaway polarization. This is because a given individual is more strongly ``anchored'' in their opinion when they have longer memories of their past expression.

\begin{figure}[!h]
    \centering
    \includegraphics[width=\linewidth]{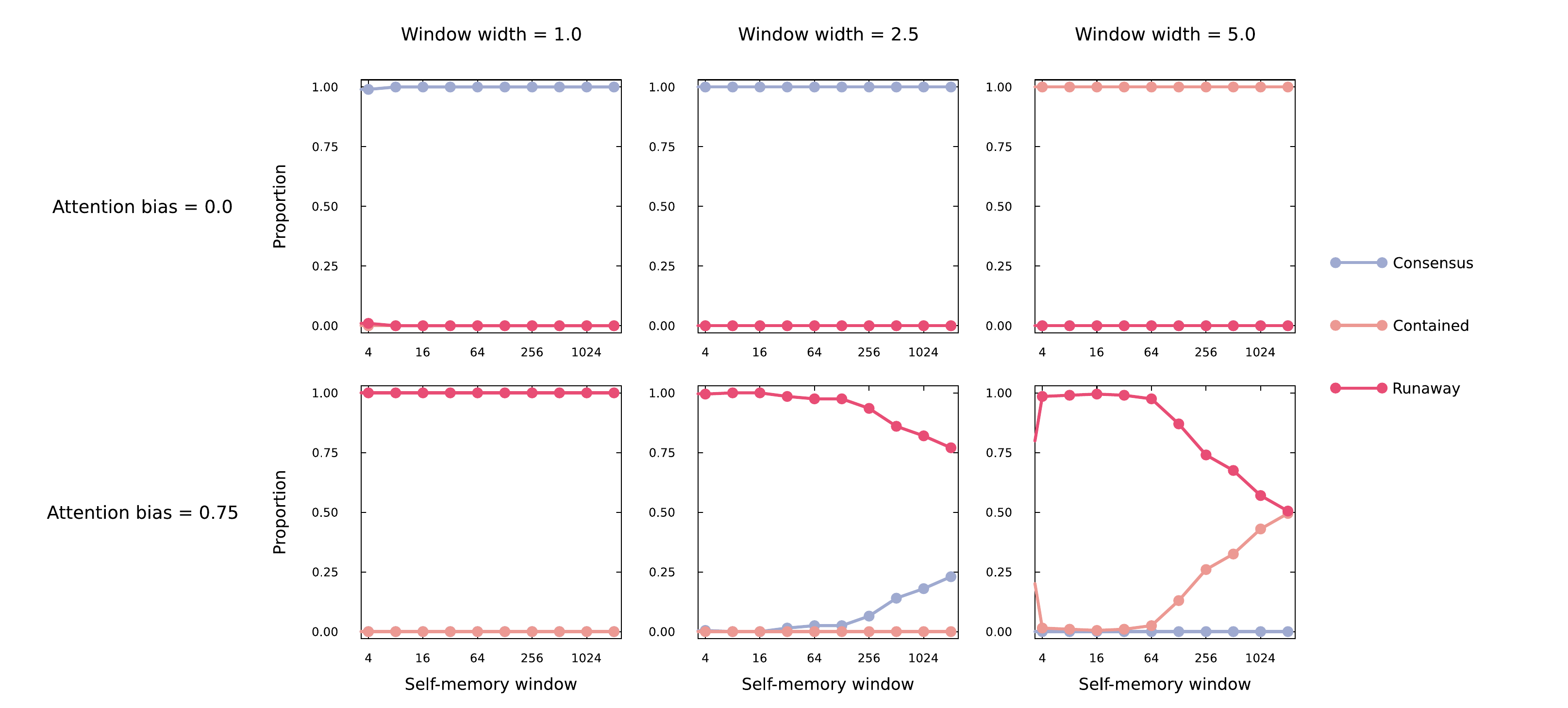}
    \caption*{\textbf{Figure S12}. The proportion of consensus, constrained polarization, and runaway polarization, for high and low values of Overton window width and Attention bias, varying the number of self-memories $s$ in the simulation. Simulations were run for the equivalent of 50 years (10 million attitude updates) with $N=100$ individuals, $n=2$ attitude dimensions, $D=2$ identity dimensions, $m=50$ memories of others, and relative attitude update step size of $r=0.2$. Utility parameters consist of self weight $\lambda_{\text{self}}=0.4$, an in-group pull of $\lambda_{\text{pull}}=0.4$, soft Overton cost of $c_S=0.5$ and hard Overton cost of $c_H=0.5$.}
\end{figure}

\newpage

Similarly, systematically increasing the number of ``social'' memories retained by an individual, $m$ (Figure S13) tends to result in a reduction in runaway polarization. This is similar to decreasing the number of interlocutors, $N$ (Figure S11). Longer social memory results in a more stable perception of in- and out-group opinions, and thus less opportunity to ``drift'' outside the hard Overton window.

\begin{figure}[!h]
    \centering
    \includegraphics[width=\linewidth]{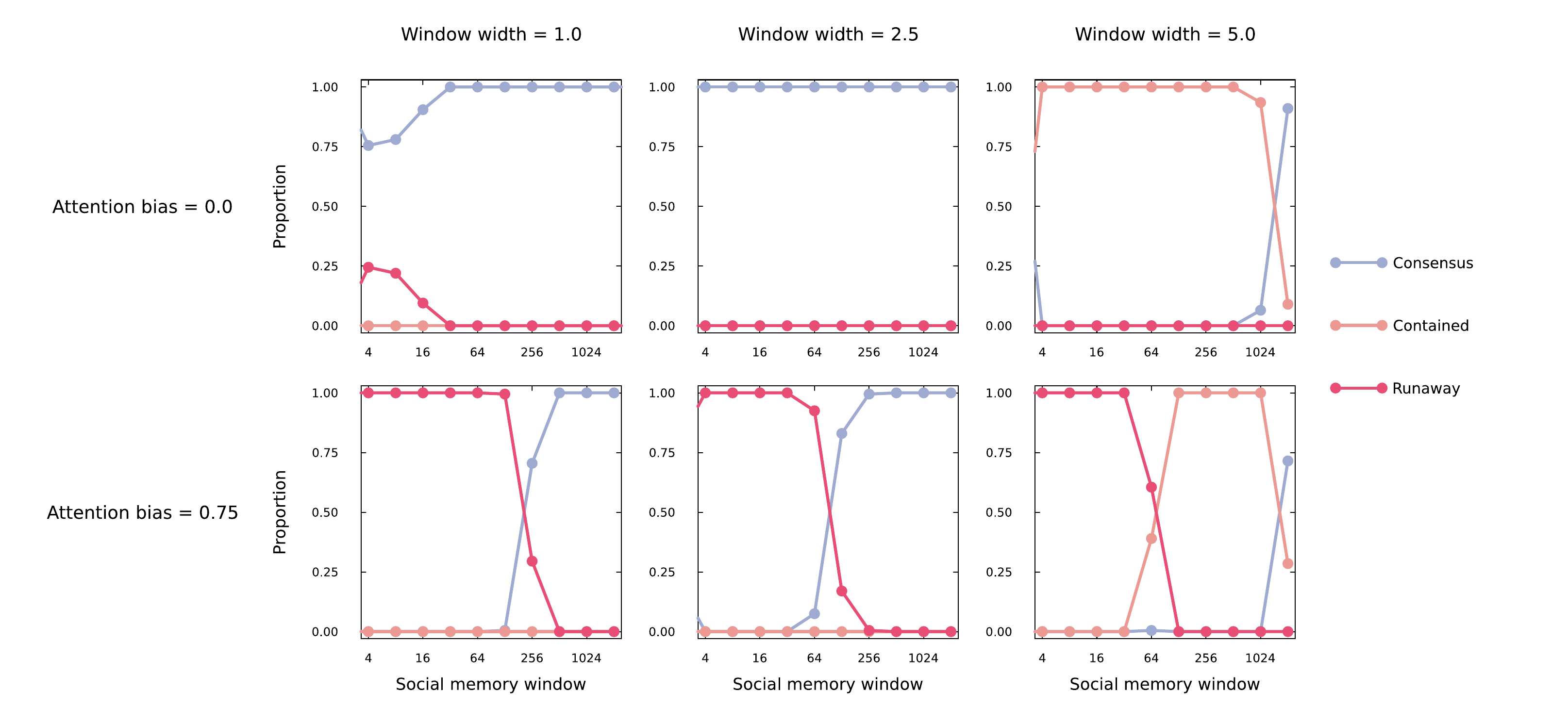}
    \caption*{\textbf{Figure S13}. The proportion of consensus, constrained polarization, and runaway polarization, for high and low values of Overton window width and Attention bias, varying the number of memories of others kept by individuals $m$ in the simulation. Simulations were run for the equivalent of 50 years (10 million attitude updates) with $N=100$ individuals, $n=2$ attitude dimensions, $D=2$ identity dimensions, $s=25$ self-memories and relative attitude update step size of $r=0.2$. Utility parameters consist of self weight  $\lambda_{\text{self}}=0.4$, an in-group pull of $\lambda_{\text{pull}}=0.4$, soft Overton cost of $c_S=0.5$ and hard Overton cost of $c_H=0.5$.}
\end{figure}

\newpage

\subsection{Interventions}

Finally we further varied the parameters associated with interventions, inlcuding their timing, strength and interaction with utility parameters.

\subsubsection{Soft and hard Overton window cost}

First we co-varied the hard Overton window cost, $c_H$, and in-group pull $\lambda_{\text{pull}}$ for high and low attention bias (Figure S14). We see that as $\lambda_{\text{pull}}$ increases, the value of $c_H$ required to prevent runaway polarization decreases.

\begin{figure}[!h]
    \centering
    \includegraphics[width=\linewidth]{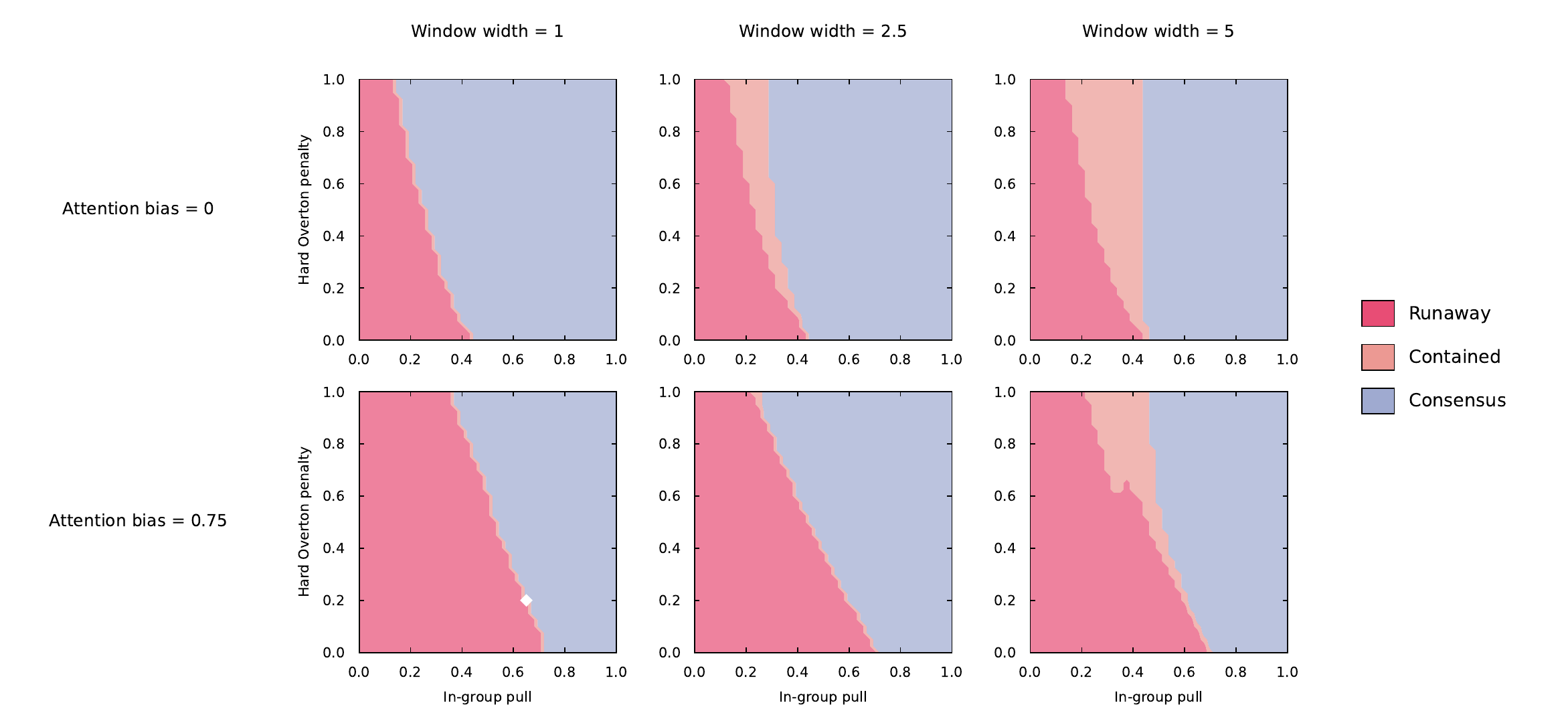}
    \caption*{\textbf{Figure S14}. The dominant outcome state across values of in-group pull $\lambda_{\text{pull}}$ and hard Overton cost $c_H$ for high and low values of attention bias and the Overton window width. Simulations were run for the equivalent of 50 years (10 million attitude updates) with $N=100$ individuals, $n=2$ attitude dimensions, $D=2$ identity dimensions, $m=50$ memories of others, $s=25$ self-memories and attitude update step size of $r=0.2$. Soft Overton cost is $c_S=0.5$. A given set of parameters is classified as a particular state if they result in that classification at least 50\% of the time over 200 replicate simulations with initial conditions randomized withing the hard Overton window. White regions corresponds to cases where no outcome state meets this threshold.}
\end{figure}

\newpage

Next we co-varied the soft Overton window cost, $c_S$, and in-group pull $\lambda_{\text{pull}}$ for high and low attention bias (Figure S15). We see a very similar pattern as for varying $c_H$, i.e. as $\lambda_{\text{pull}}$ increases, the value of $c_S$ required to prevent runaway polarization decreases.

\begin{figure}[!h]
    \centering
    \includegraphics[width=\linewidth]{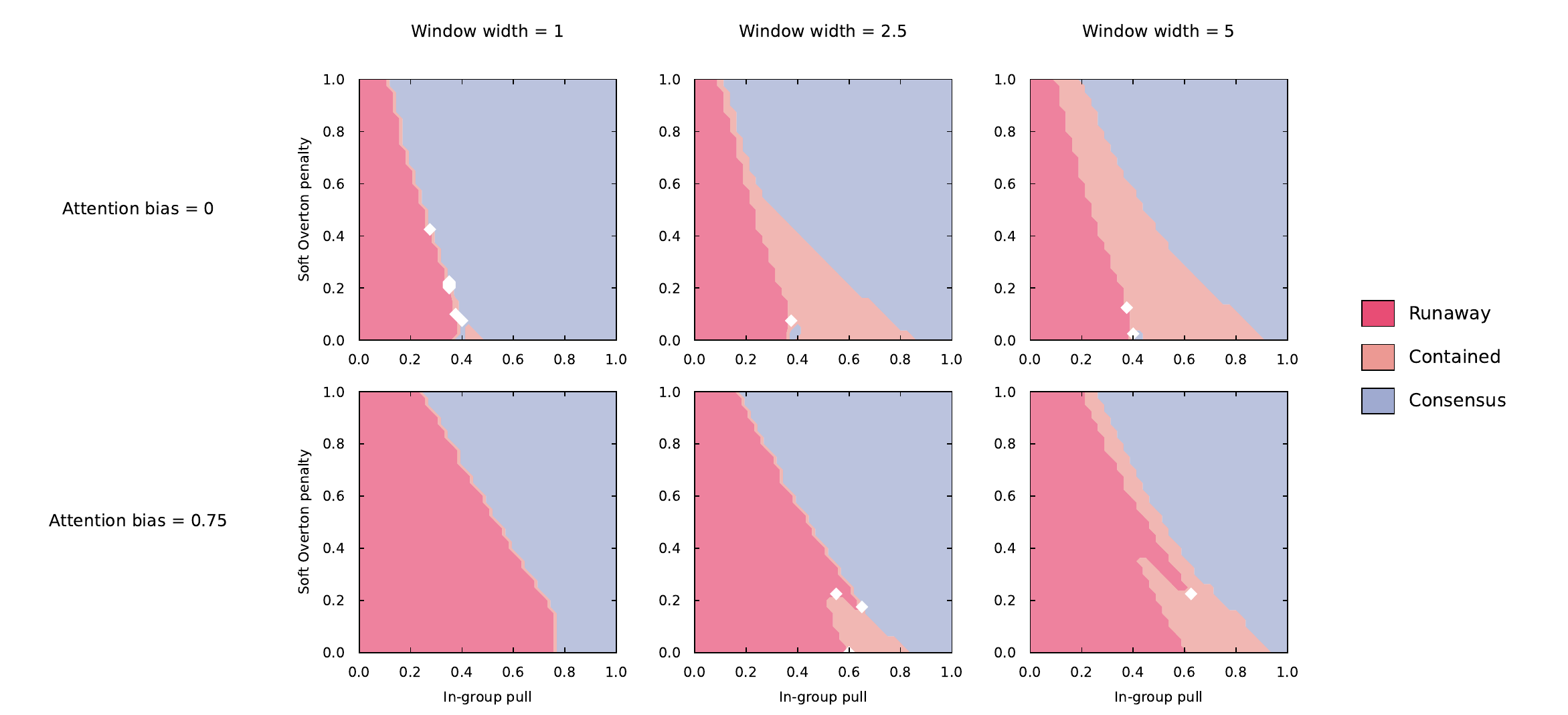}
    \caption*{\textbf{Figure S15}. The dominant outcome state across values of in-group pull $\lambda_{\text{pull}}$ and soft Overton cost $c_S$ for high and low values of attention bias and the Overton window width. Simulations were run for the equivalent of 50 years (10 million attitude updates) with $N=100$ individuals, $n=2$ attitude dimensions, $D=2$ identity dimensions, $m=50$ memories of others, $s=25$ self-memories and attitude update step size of $r=0.2$. Hard Overton cost is $c_H=0.5$. A given set of parameters is classified as a particular state if they result in that classification at least 50\% of the time over 200 replicate simulations with initial conditions randomized withing the hard Overton window. White regions corresponds to cases where no outcome state meets this threshold.}
\end{figure}

\newpage

\subsubsection{Intervention efficacy}

In Figure S16 we compare the outcome state with and without a fixed time intervention over a range of values for $w_H$ and $\beta$. Note these results are a more detailed breakdown of those summarized in Main Text Figure 4.

\begin{figure}[!h]
    \centering
    \includegraphics[width=0.8\linewidth]{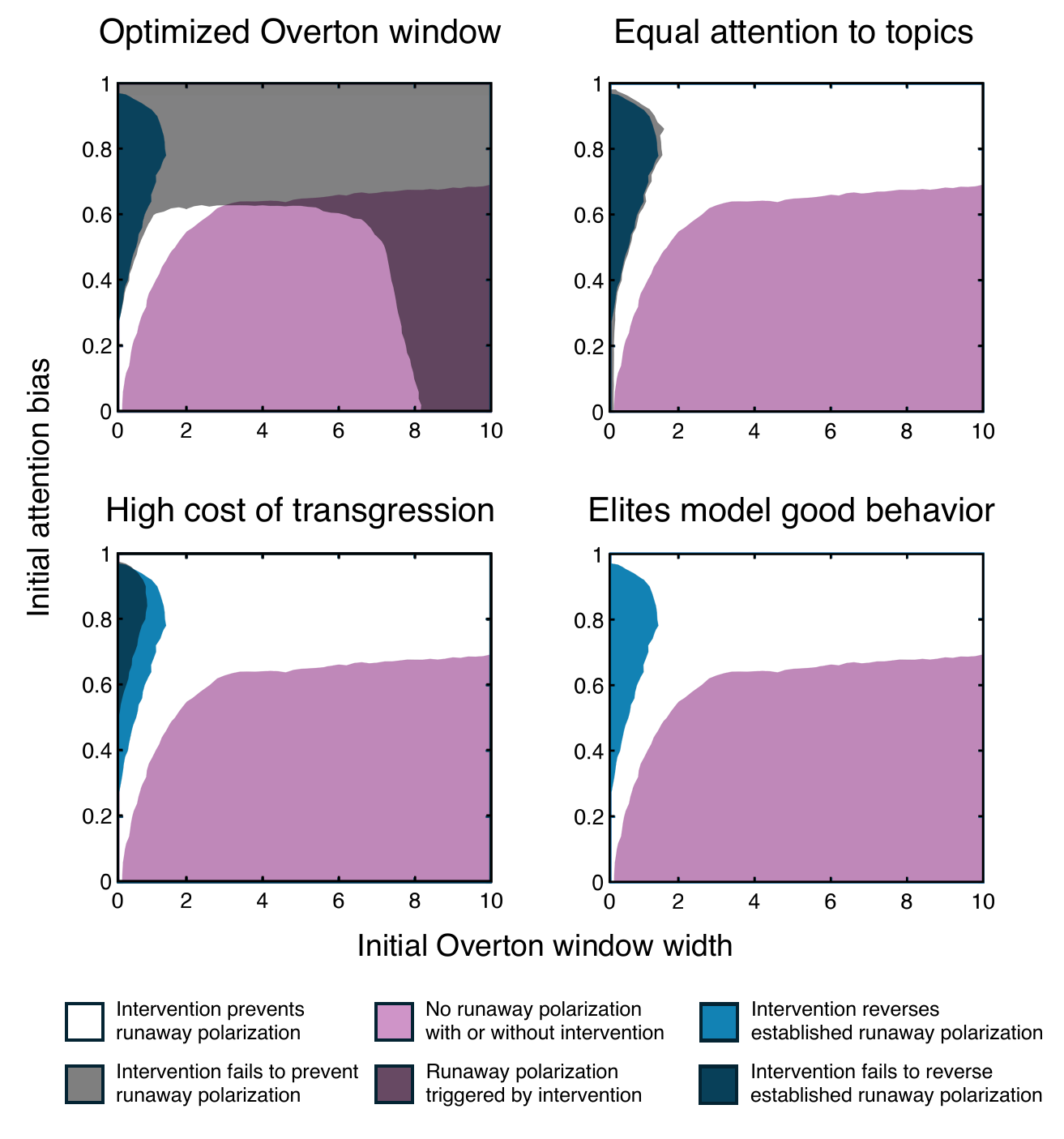}
    \caption*{\textbf{Figure S16}.We calculated the average efficacy of the four interventions described in the main text across hard Overton window widths varying from 0.05 to 10 and attention bias varying from 0 to 1. For each of the four panels, the color describes the state of the system with and without the intervention, as summarized in the key. Simulations are run for the equivalent of 50 years (10 million attitude updates) with interventions implemented after 2 years (500 000 attitude updates). Simulations are run with $N=100$ individuals, $n=2$ attitude dimensions, $D=2$ identity dimensions, $m=50$ memories of others, $s=25$ self-memories and attitude update step size of $r=0.2$. Self weight is $\lambda_{\text{self}}=0.4$ and in-group pull is $\lambda_{\text{self}}=0.4$. Soft Overton cost is $c_S=0.5$ and hard Overton cost is $c_H=0.5$. The cost hike intervention hikes the cost of transgressing the Overton window to $c_H=50$ (100 fold), the window optimization intervention sets the Overton window width to 3.0, and the weight of the ingroup pull of moderate elites is set to 1.0 for the moderate elites intervention (see Methods). The attention balance intervention sets $\alpha=0$.}
    \label{fig:placeholder}
\end{figure}

\newpage

In Figure S17 we show the effect of intervention lag-time on the efficacy of interventions at reversing runaway polarization. This is the same data as used to generate Main Text Figure 5a, but we also show results for additional interventions and for $D=1$ and $D=10$ identity dimensions, as well as for $D=2$. We see in particular that the attention balance intervention can be effective if implemented very quickly after runaway polarization takes hold, and when the number of identity dimensions is high.

\begin{figure}[!h]
    \centering
    \includegraphics[width=0.8\linewidth]{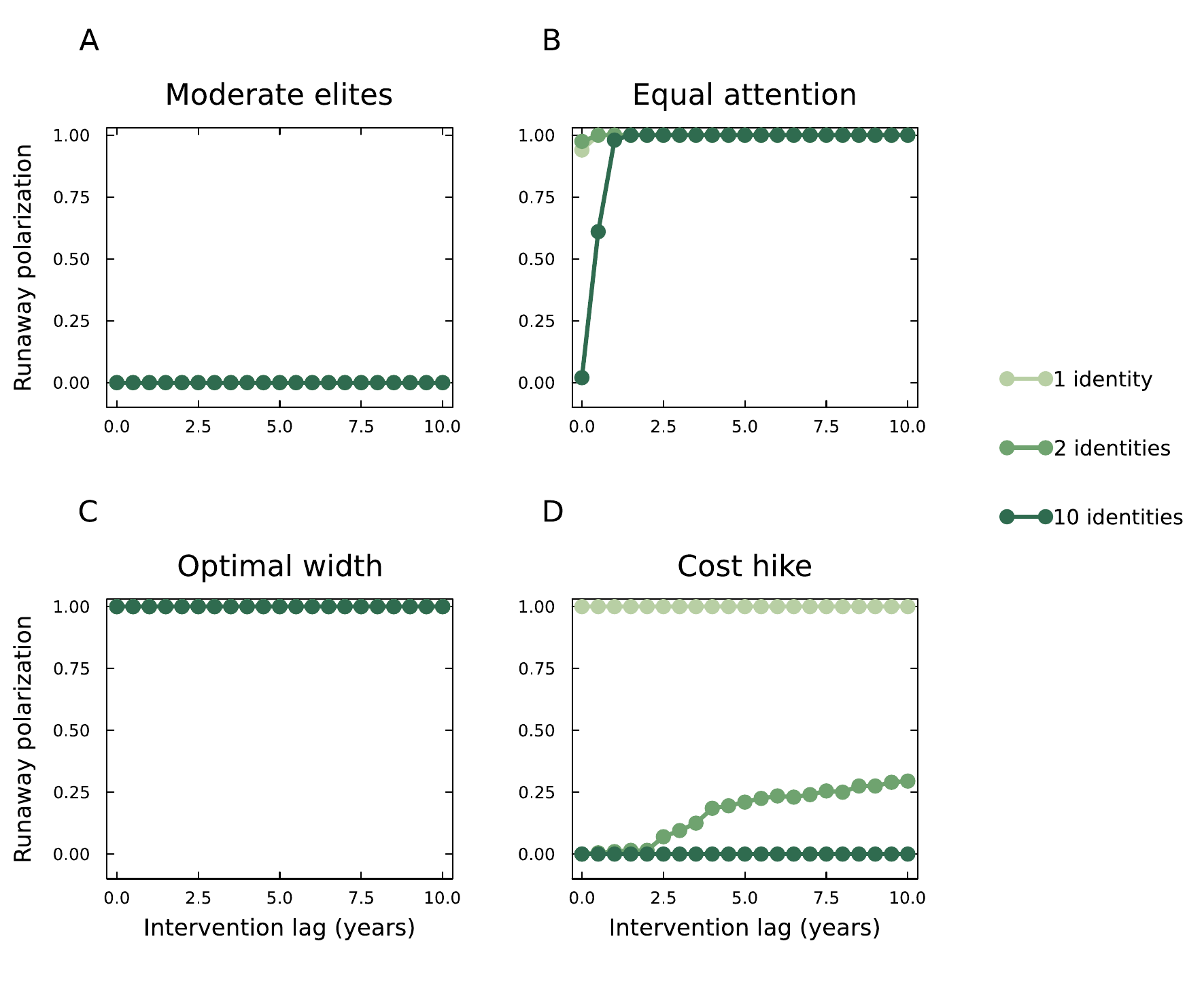}
    \caption*{\textbf{Figure S17}. Panels display runaway polarization for our four different interventions across different time lags from the onset of polarization to intervention implementation. Each value is an average over 200 replicates. Simulations were run for the equivalent of 50 years (10 million attitude updates) with $N=100$ individuals, $n=2$ attitude dimensions, $D=2$ identity dimensions, $m=50$ memories of others, $s=25$ self-memories and relative attitude update step size of $r=0.2$. Utility parameters consist of self weight $\lambda_{\text{self}}=0.4$, an out-group pull of $\lambda_{\text{pull}}=0.4$, soft Overton cost of $c_S=0.5$, hard Overton cost of $c_H=0.5$, attention bias of $\alpha=0.75$, and Overton window width of $w_H=1.0$.}
    \label{fig:placeholder}
\end{figure}

\newpage

In Figure S18 we explore the effect of the number of identity dimensions on interventions implemented at a fixed time. We see as that only the cost hike intervention is sensitive to the number of identity dimensions, and is more effective when identites are more complex.

\begin{figure}[!h]
    \centering
    \includegraphics[width=0.6\linewidth]{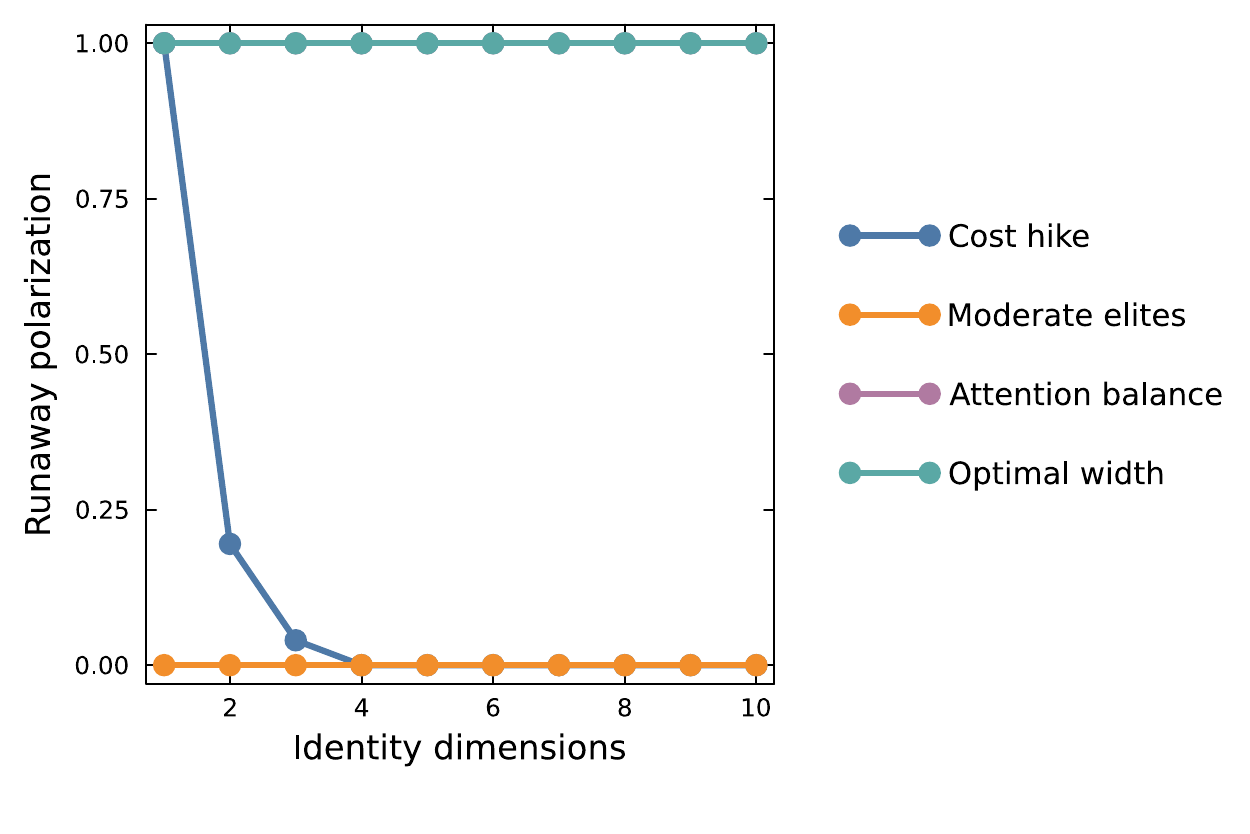}
    \caption*{\textbf{Figure S18}. Runaway polarization proportion for our four different interventions across different numbers of identity dimensions. Simulations were run for the equivalent of 50 years (10 million attitude updates) with $N=100$ individuals, $n=2$ attitude dimensions, $m=50$ memories of others, $s=25$ self-memories and relative attitude update step size of $r=0.2$. Utility parameters consist of self weight $\lambda_{\text{self}}=0.4$, an out-group pull of $\lambda_{\text{pull}}=0.4$, soft Overton cost of $c_S=0.5$, hard Overton cost of $c_H=0.5$, attention bias of $\alpha=0.75$, and Overton window width of $w_H=1.0$. Interventions were implemented at the equivalent of 4 years after simulation initialization.}
    \label{fig:placeholder}
\end{figure}

\newpage

\subsubsection{Intervention strength}

Finally we explore the impact of intervention strength, for the two most effective interventions identified in the main text i.e. cost hike and amplification of moderate elites. Figure S19 shows the impact increasing the pull of moderate elites for different hard Overton window widths, and different levels of attention bias. We see that in all cases, there is a sharp transition above which moderate elites become highly effective.

\begin{figure}[!h]
    \centering
    \includegraphics[width=\linewidth]{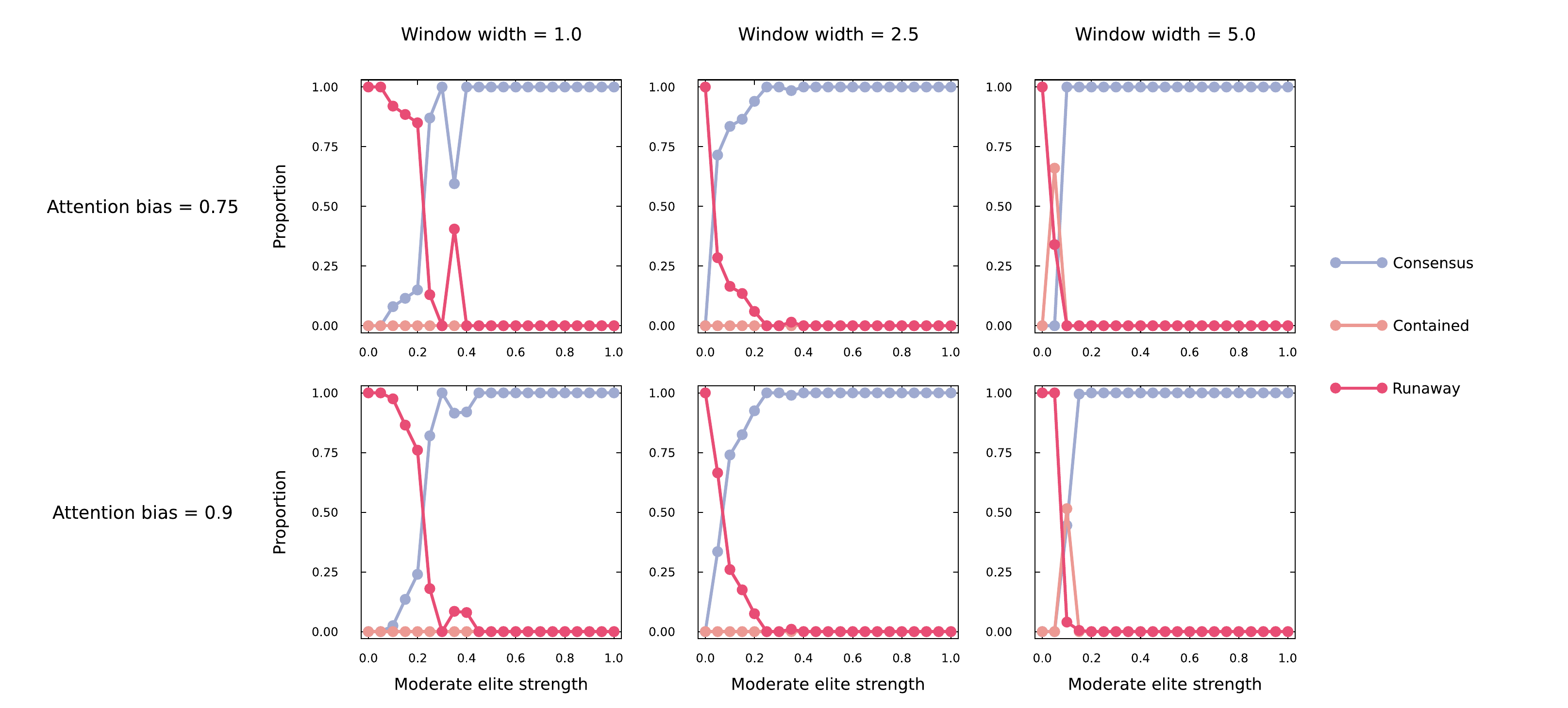}
    \caption*{\textbf{Figure S19}. The proportion of consensus, constrained polarization, and runaway polarization, for high and low values of Overton window width and Attention bias, varying the strength of a Moderate elites intervention implemented at a fixed time after model initialization (500 thousand attitude updates). Simulations were run for the equivalent of 50 years (10 million attitude updates) with $N=100$ individuals, $n=2$ attitude dimensions, $D=2$ identity dimensions, $m=50$ memories of others, $s=25$ self-memories and relative attitude update step size of $r=0.2$. Utility parameters consist of self weight  $\lambda_{\text{self}}=0.4$, an in-group pull of $\lambda_{\text{pull}}=0.4$, soft Overton cost of $c_S=0.5$ and hard Overton cost of $c_H=0.5$.}
\end{figure}

\newpage

Similar results for the cost hike intervention (Figure S20) show a similar story, although the transition is less sharp than for the moderate elites, with intermediate levels of runaway polarization often persisting even when the cost hike is very large.

\begin{figure}[!h]
    \centering
    \includegraphics[width=\linewidth]{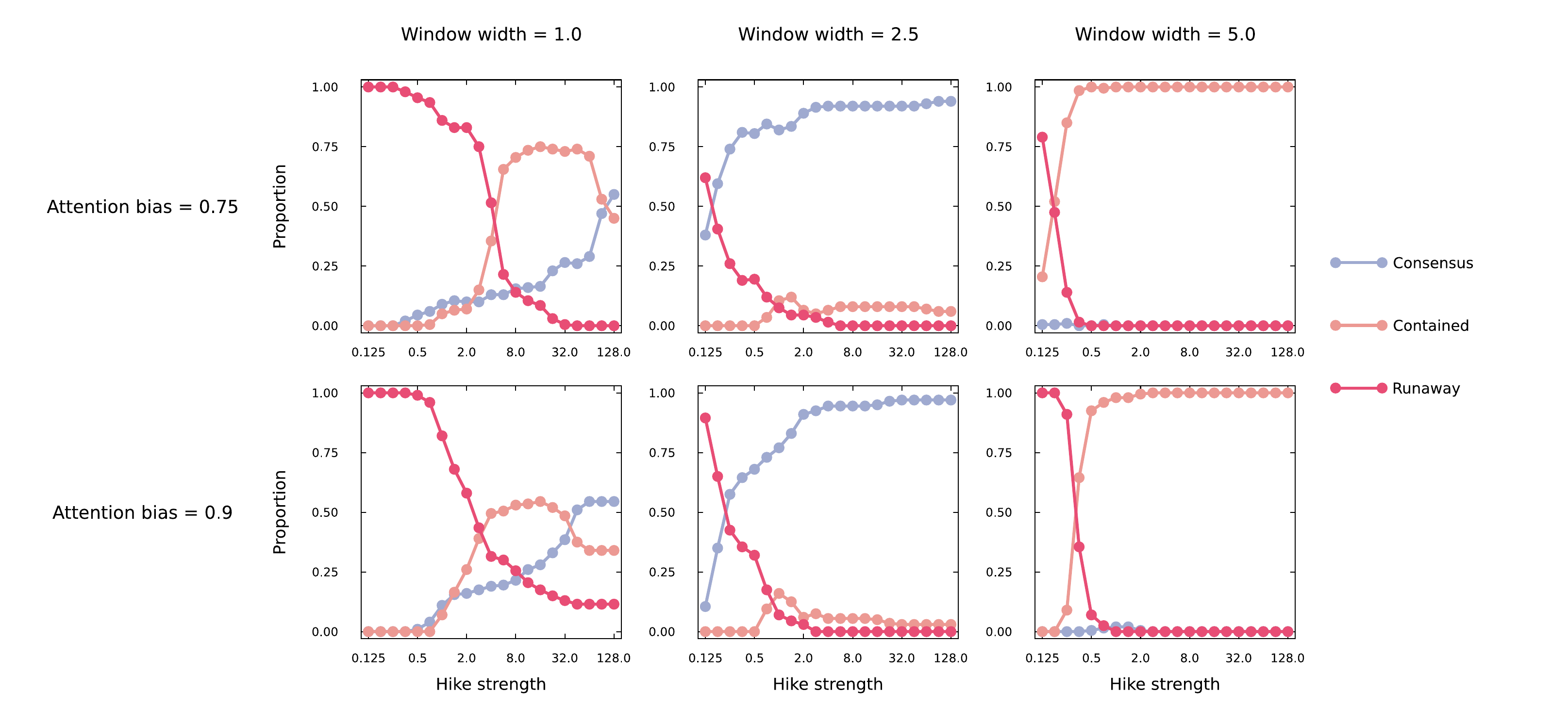}
    \caption*{\textbf{Figure S20}. The proportion of consensus, constrained polarization, and runaway polarization, for high and low values of Overton window width and Attention bias, varying the strength of a Cost hike intervention implemented at a fixed time after model initialization (500 thousand attitude updates). Simulations were run for the equivalent of 50 years (10 million attitude updates) with $N=100$ individuals, $n=2$ attitude dimensions, $D=2$ identity dimensions, $m=50$ memories of others, $s=25$ self-memories and relative attitude update step size of $r=0.2$. Utility parameters consist of self weight  $\lambda_{\text{self}}=0.4$, an in-group pull of $\lambda_{\text{pull}}=0.4$, soft Overton cost of $c_S=0.5$ and hard Overton cost of $c_H=0.5$.}
\end{figure}

\newpage

\section{Code}

The code for our model is available via a github repository \href{https://github.com/LeonKlingborg/preventing-polarized-discourse}{\textcolor{blue}{here}}.


\end{document}